# Author's Post-Print (final draft post-refereeing)

# Online Force-Directed Algorithms for Visualization of Dynamic Graphs

Se-Hang Cheong[a,*], Yain-Whar Si[a], Raymond K. Wong[b]

[a]Department of Computer and Information Science, University of Macau
[b]School of Computer Science and Engineering, University of New South Wales

**Abstract**

Force-directed (FD) algorithms can be used to explore relationships in social networks, visualize money markets, and analyze transaction networks. However, FD algorithms are mainly designed for visualizing static graphs in which the topology of the networks remains constant throughout the calculation. In contrast to static graphs, nodes and edges in dynamic graphs can be added or removed as time progresses. In these situations, existing FD algorithms do not scale well, since any changes in the topology will trigger these algorithms to completely restart the entire computation. To alleviate this problem, we propose a design and implementation of five online FD algorithms to visualize dynamic graphs while maintaining their native force models. The online FD algorithms developed in this paper are able to reuse the force models of existing FD algorithms without significant modifications. To evaluate the effectiveness of the proposed approach, online FD algorithms are compared against static FD algorithms for visualizing dynamic graphs. Experimental results show that among the five algorithms evaluated, the online FD algorithm achieves the best number of edge crossings and the standard deviation of edge lengths for visualizing dynamic graphs.

## 1. Introduction

Force-directed (FD) algorithms are considered as one of the main techniques for visualization of graphs. These algorithms include FR [1], LinLog [2], KK [3], DH [4], and FA2 [5] algorithms. The main purpose of these algorithms is to visualize static graphs in which the topology of the networks remains constant throughout the calculation. FD algorithms can also produce visualizations purely based on the topology of a graph and do not require additional information to generate a visualization. FD algorithms are well suited for the visualization of graphs since they can be easily adapted for different network topologies. For example, the Fruchterman Reingold (FR) algorithm [1], the Kamada-Kawai (KK) algorithm [3], ForceAtlas2 (FA2) algorithm [5], and Davidson Harel (DH) algorithm [4] were used for edge crossing reduction and planar graph drawing. The Linlog algorithm [2] was also adopted for visualization of graph clustering, etc. We denote these algorithms as static force-directed (SFD) algorithm since their main objective is to generate visualization of static graphs.

---
*Corresponding author
 *Email addresses:* dit.dhc@lostcity-studio.com (Se-Hang Cheong), fstasp@umac.mo (Yain-Whar Si), wong@cse.unsw.edu.au (Raymond K. Wong)

However, in many applications, constantly changing information may be modeled as a sequence of static graphs in which each graph is associated with a timestamp. These graphs are commonly denoted as dynamic graphs in data analytics and visualization communities. In these applications, visualization of dynamic graphs is performed as a crucial step in discovering meaningful information [6]. Visualization of dynamic graphs allows analysts to observe any changes in the graphs over time to find hidden relationships and patterns. When processing large-scale dynamic graphs, the "big" picture (i.e. the layout outline of graph) of interesting points requires to be recomputed continuously. Such computation needs to be efficiently performed so that outliers, transitions and trends can be clearly visible in the "big" picture. [7].

Since early nineties, several visualization of dynamic graphs have been proposed [8]. These visualizations of dynamic graphs include finding relationships and communities in social networks [9, 10], visualizing metabolic networks and gene structures in bioinformatics [11, 12], and money market visualization of interbank and large transaction networks [13, 14], etc. Visualization of dynamic graphs are also useful in real-time monitoring of supply chains. For example, monitoring flow of goods in warehousing, transportation, plants and other infrastructures in supply chain can be achieved by visualizing dynamic graphs (or networks) which contains nodes and edges representing entities and connections respectively. Monitoring and optimizing flows in these networks may need to be performed in real-time by supply chain managers, because adjustments are made when the conditions in these networks change [15]. However, methods that apply to static graphs are insufficient for dynamic graphs, since nodes and edges may appear or disappear over time in dynamic graphs.

Several criteria for improving the stability and readability of the visualization of dynamic graphs are reported in literature [16, 17]. For example, in *mental map preservation*, the placement of nodes and edges between adjacent timeslices was recommended to change as little as possible [18, 19, 20]. Another approach called *small multiples* is also commonly used for visualization in which multiple timeslices are shown on the screen simultaneously. However, small multiples and mental map preservation are not useful in all scenarios. Specifically, small multiples are difficult in distinguishing how a graph evolves [8], due to too many similar layouts presented on a screen. Mental map preservation may not be suitable for active and large-scale dynamic graphs because a large volume of nodes and edges may change rapidly.

Drawing layouts by using Node-and-Link layouts [15] is an important step in the visualization of dynamic graphs. Online algorithms are often used to produce a Node-and-Link layout of a dynamic graph with a series of timeslices [21]. Updates in nodes and edges can be instantly reflected by these online algorithms. In addition, online algorithms are also used for learning similarity between pairs of nodes in large scale images [22] and segmenting time series for mining transactions [23]. Therefore, online algorithms for dynamic graph visualization should be designed with mechanisms to compute a complete picture rather than looking at the details of each static graph. Moreover, the online algorithms should be flexible to deal with any changes in the underlying networks. Finally, these algorithms should be able to handle large and active dynamic graphs.

To achieve these objectives, in this paper, we propose a systematic approach to design and implement five online force-directed (OFD) algorithms for visualization of dynamic graphs using Node-and-Link layouts. In our approach, OFD algorithms are extended from existing SFD algorithms by integrating with a set of common modules which are designed to address the key issues such as handling of continual updates in graph topology, managing memory and IO handling for processing large scale topology, and reducing overall re-calculation to boost the algorithm performance. One of the main advantages of the proposed design is that, the OFD algorithms developed in this paper are able to reuse the force models of existing FD algorithms without significant modifications. Therefore, OFD algorithms inherit the properties of the original FD algorithms and allow users to examine the underlying strength and weakness of these OFD algorithms.

However, there are challenges and limitations in adopting SFD algorithms for visualization of dynamic graphs. For instance, complexity is one of major drawbacks since a large number of iterations are often needed for layout drawing [24]. In this paper, we show that the proposed OFD algorithms can significantly lower the computational cost for visualization of large dynamic graphs. Since the OFD algorithms can process large dynamic graphs, they are useful for visualization tasks in transaction monitoring, stock trading analysis, and online payment analysis. The contributions of this paper are summarized as follows:

- We describe a comprehensive analysis of existing five SFD algorithms. Based on the analysis, a common architecture of SFD algorithms is presented.

- We present a novel systematic approach for developing OFD algorithms based on the proposed, common architecture.

- Based on the systematic approach, a design and implementation of five OFD algorithms for dynamic graph visualization are presented. These algorithms achieve key properties including reusability, scalability, and compatibility.

The rest of this article is organized as follows. In Section 2, we review related work on force-directed algorithms for visualizing dynamic graphs. In Section 3, we discuss the idea behind the design and implementation of OFD algorithms for visualization of dynamic graphs. In Section 4, we evaluate the performance of OFD algorithms for visualizing dynamic graphs from communication, transaction and social networks. In Section 5, we conclude the paper.

## 2. Related Work

Transaction monitoring and social network visualizations are some of the well-known visualization of dynamic graphs. For instance, analyzing the purchasing pattern over time is one of typical visualizations in transaction graphs which can be formulated as a problem of dynamic graph analysis [15]. A system to monitor interbank payment transactions by analyzing payment patterns over several days is also proposed in [25]. Heijmans et al. [25] showed that the payment transactions can be visualized in dynamic graphs to support exploratory research and provide early warning information for interbank payments.

Although SFD algorithms are widely used for the layout drawing of static graphs, several approaches have been proposed for visualization of historical data in dynamic graphs. In these approaches, SFD algorithms are used to animate a sequence of graphs or display selected nodes side-by-side in small multiples [26]. For example, visualization of transaction patterns of Bitcoin is proposed in [27] based on the FA2 algorithm [5]. Visualization of transactions flow of Bitcoin based on a SFD algorithm is also proposed in [28]. In their approach, transaction flows of Bitcoin graphs are visualized in terms of temporal properties and the amount of Bitcoin spent. These approaches use a batch processing provided in SFD algorithms. In addition, transactions are considered as off-line data in these approaches and the information of nodes and edges for the dynamic graphs are extracted from historical data. Although these studies can be used to analyze dynamic graphs based on the historical data, they are quite limited in the scope of visualization of dynamic graphs.

Visualization of dynamic graphs are challenging since dynamic graphs are constantly changing overtime. Besides, the information on how and when the dynamic graph might change over time is often unavailable since nodes and edges could appear or disappear in an ad hoc manner and the volume and arrival rate of incoming data are unpredictable. In addition, small multiple and mental map preservation may not always be suitable for active and large dynamic graphs which already contain too much details. Therefore, methods that apply to static graphs are not sufficient for the visualization of dynamic graphs. In addition, managing how the temporal properties should be presented in a layout drawing that is a also crucial problem in dynamic graphs. In previous studies, there are two major approaches often used for the representation of time in dynamic graphs: timeline (time-to-space) and animation (time-to-time). In timeline approach, the time dimension was mapped to a spatial dimension. Matrix-based layouts such as Adjacency Matrix Layout, Parallel Coordinates Layout are often used to display sequential graphs on a timeline [29]. Node-and-Link layouts [15] are often used for the animation approach [30] where each layout represents the state (snapshot) of dynamic graphs at a given time.

Furthermore, SFD algorithms are also used for the representation of time in dynamic graphs. Erten et al. [31] applied a SFD algorithm to animate dynamic graphs over the series of timeslices. SFD algorithms with GPU acceleration is also proposed by [16]. In their approach, they used a weighted function to preserve the mental map for dynamic graphs. Besides, there are also studies for the visualization without the use of Node-and-Link layouts. The visualization method to omit the details of a graph is proposed by [32]. Their approach visualizes graphs by a Bipartite layout. The objective of their approach is to view a sequence of dynamic graphs on a screen as long as possible. Moreover, Brandes and Wagner [33] proposed a SFD algorithm for the visualization of dynamic and undirected graphs using a Bayesian paradigm. Bipartite layout and Bayesian paradigm are useful in the visualization of high dimensional data to distinguish the difference in a specific attribute. Node-and-Link layouts are commonly used in graph drawing and they are especially useful for analyzing the relationship among the nodes [34]. Therefore, in this paper, we consider Node-and-Link layouts for the visualization of dynamic graphs.

SFD algorithms are suitable for computing an overall layout when details of the entire graph is less important. Therefore, these algorithms can be easily adopted for dynamic graph visualization tasks. In this paper, we address acceleration and heuristic techniques for visualizing dynamic graphs in OFD algorithms. Specifically, in our approach, OFD algorithms are extended from existing SFD algorithms by (1) integrating with a set of common modules which are designed to address the key issues such as handling of continual updates in graph topologies, (2) managing memory and IO handling for processing large scale topology, and (3) reducing overall re-calculation to boost the algorithm's performance. Various force models have been proposed for SFD algorithms in recent years. From our analysis, we find that although these force models are inherently different, they share several common components. Based on the above, we propose a design of OFD algorithms where SFD algorithms can be easily extended without significant modifications. The details of SFD and OFD algorithms are further analyzed in the following section.

## 3. Online Force-directed Algorithms for Visualization of Dynamic Graphs

Static Force-directed (SFD) algorithms have been widely adopted for visualization of static graphs. Majority of the SFD algorithms have common objectives such as minimizing the edge crossing and producing planer graphs. Moreover, by using proper heuristics, SFD algorithms are shown to achieve desired stability and readability properties for generating the layouts of static graphs. Therefore, SFD algorithms are selected as a candidate among other visualization techniques for visualization of dynamic graphs. In this paper, we propose a systematic approach for designing OFD algorithms for visualization of dynamic graphs. To better explain the idea of the proposed design, the definition of online dynamic graphs is described in Section 3.1. The comprehensive analysis of existing SFD algorithms is described in Section 3.2. The overall approach is detailed in Section 3.3.

*3.1. Definitions*

A dynamic graph *D* is modeled as a sequence of static graphs *G* as defined in Equation 1. In this paper, each static graph *G* is a directed graph and associated with a time stamp.

$$D := (G_1, ...., G_n) \qquad (1)$$

Here, we define

$$G := (V, E) \qquad (2)$$

be a static graph with node set *V* and edge set *E* where nodes $v \in V$ and edges $e \in E$. An edge

$$e = (u, v), u \neq v \qquad (3)$$

is a connection between two nodes. The layout of a static graph maps the nodes into two-dimensional positions on the canvas. $p_v$ is the position of a node *v* in the layout ($p_v : V \rightarrow R^n$). Moreover, we define function *dG*(*u, v*) which returns the Euclidean distance between nodes *u* and *v*. Function *degree*(*u*) returns the total in and out degrees of the node *u*.

*3.2. Analysis of Existing Force-directed Algorithms*

In this section, we aim to provide a generalized view of the underlying architecture of each SFD algorithm under consideration. Five SFD algorithms are analyzed in this paper. They are FR [1], LinLog [2], KK [3], DH [4], and FA2 [5] algorithms. By understanding the underlying design of each algorithm, we can further enhance each algorithm in a consistent way to achieve our main objective, that is to visualize dynamic graphs.

SFD algorithms are designed to process nodes (vertex) and edges (link). In addition, graph and layout are two fundamental data structures used in the SFD algorithms. Topology information such as nodes and their connections (i.e. edges) are part of a graph. Layout is used in the SFD algorithms to determine the locations where each of its nodes is going to be placed. The information of layout is also used for graph drawing and determine the size of the canvas, etc.

In our analysis, nodes, edges, graphs and layouts are considered as fundamental data structures, whilst other components are considered as algorithm specific. We also categorize the tasks of SFD algorithms into common and specific tasks. Common tasks include tasks which exist in most SFD algorithms such as functions for canvas size determination, force constant initialization, iteration determination and updating nodes' position. Specific tasks are unique tasks which are different from one SFD algorithm to another. These tasks could be used for internal data structure initialization, graph partitioning, building Octree, etc. The summary of common and specific tasks of SFD algorithms are shown in Figure 1. In this diagram, common and specific tasks are highlighted in blue and orange solid blocks respectively. According to Figure 1, we can observe that more than 50% of the tasks can be identified as common tasks in the SFD algorithms analyzed.

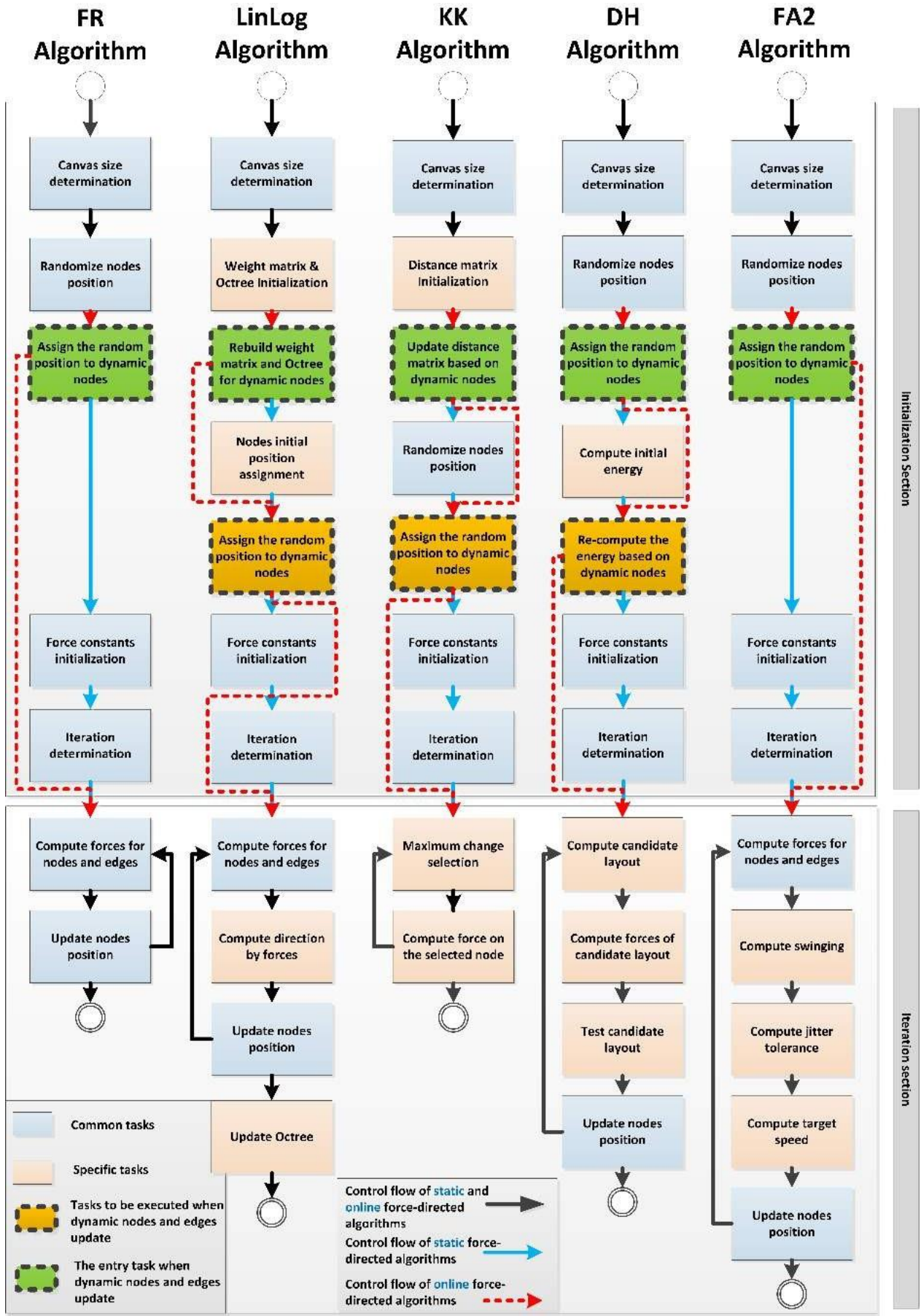

Figure 1: Common and specific tasks of SFD and OFD algorithms.

To further understand the common and specific tasks of SFD algorithms, we have also created a set of detailed Data Flow Diagrams (DFD) for each SFD algorithm from Figure 1. We then implemented these SFD algorithms according to the DFD diagrams. Due to the page limitation, the DFD diagrams are published in authors' homepage [1]. Based on these DFD diagrams, we further divide the control flow of SFD algorithms into two sections:

- *Initialization section:* Canvas size, constants, nodes position and data structure initialization are set in the initialization section. Besides, majority of SFD algorithms use random positions as initial positions of nodes. In Linlog algorithm, the initial position of nodes is based on the weight matrix of Octree partition. In DH algorithm, the initial energy of the graph is calculated based on the nodes with randomized position. In KK and LinLog algorithms, the distance matrix and Octree are constructed in the initialization section.

- *Iteration section:* SFD algorithms repeat the processes in the iteration section until a "good" layout is produced. Most of SFD algorithms update the position of nodes after the force computation except the KK algorithm, which selected one node per iteration and update the position of the selected node. FA2 and FR algorithms are similar. However in FA2 algorithm, swinging and target speed of nodes are computed in iteration section. The higher the swinging of nodes, the slower the target speed of nodes. Moreover, only Linlog algorithm uses the direction of node pairs to compute forces. The iteration process of DH algorithm is also special because it requires at least two layouts. The initial layout is initialized in the initialization section. DH algorithm then computes new candidate layouts in the iteration section to compare with the initial layout. The layout with smaller energy will be remained for the next iteration. The layout with high energy is dropped.

*3.2.1. Transforming into Online Force-directed Algorithms*

The proposed approach to transform SFD algorithms into OFD algorithms is described in this section. In order to achieve re-usability and to inherit the properties of the original algorithms, OFD algorithms are implemented based on the force model of underlying SFD algorithms. To distinguish the tasks between SFD and OFD algorithms, we defined two special types of tasks in OFD algorithms and they are shown in Figure 1. The first type of tasks are the entry tasks when dynamic nodes and edges are updated. The second type of tasks are those tasks which are to be executed when dynamic nodes and edges are updated.

In Figure 1, black solid lines specifies the control flow of task used in both SFD and OFD algorithms. Blue solid lines specify the control flow of task which are disabled in OFD algorithms. Red dashed lines specify the control flow of task used by OFD algorithms only. OFD algorithms do not necessarily start from the scratch at the time when the dynamic graph updates. They start from the entry tasks which are depicted in green background color with dashed borders. Tasks with orange background color and dashed border specify the tasks which are only to be executed once at the time when the dynamic graph updates.

---

[1]http://eric.lostcity-studio.com/uncategorized/architecture-of-force-directed-algorithms/

Online FR, FA2 and DH algorithms have only few specific tasks according to the architecture illustrated in Figure 1. However, the online KK algorithm requires a distance matrix and online LinLog algorithm relies on weight matrix and Octree. We improved the implementation of these two algorithms. In the case of online KK algorithm, the implementation of distance matrix by an array was avoided. HashMap is also not used because all associated key pairs need to be traversed in the HashMap once a node is inserted. Iterating over HashMap is not efficient especially in large dynamic graphs. Therefore, linked-list array implementation was used for distance matrix in our implementation. In the case of online LinLog algorithm, the weight matrix should be implemented as linked-list array similar to the online KK algorithm that weights was recalculated at the time the dynamic graph updates. Moreover, the Octree used in the online LinLog algorithm was implemented in a way to support dynamic nodes insertion and deletion.

*3.3. Design of OFD Algorithms*

In this section, we detail the design of OFD algorithms. The components and corresponding control flow of the architecture are illustrated in Figure 2. Five fundamental components considered in the design of OFD algorithm for visualization of dynamic graphs are: (1) graph file input, (2) data structures, (3) force-directed algorithms, (4) online event trigger, (5) layout drawings. The description of graph file input, data structures, force-directed algorithms, online event trigger and layout drawings are described in Section 3.3.1, 3.3.2, 3.3.3, 3.3.4 and 3.3.5 respectively. Moreover, heuristics used for the node placement and the normalization of canvas scaling will be introduced in Section 3.3.6 and 3.3.7 respectively.

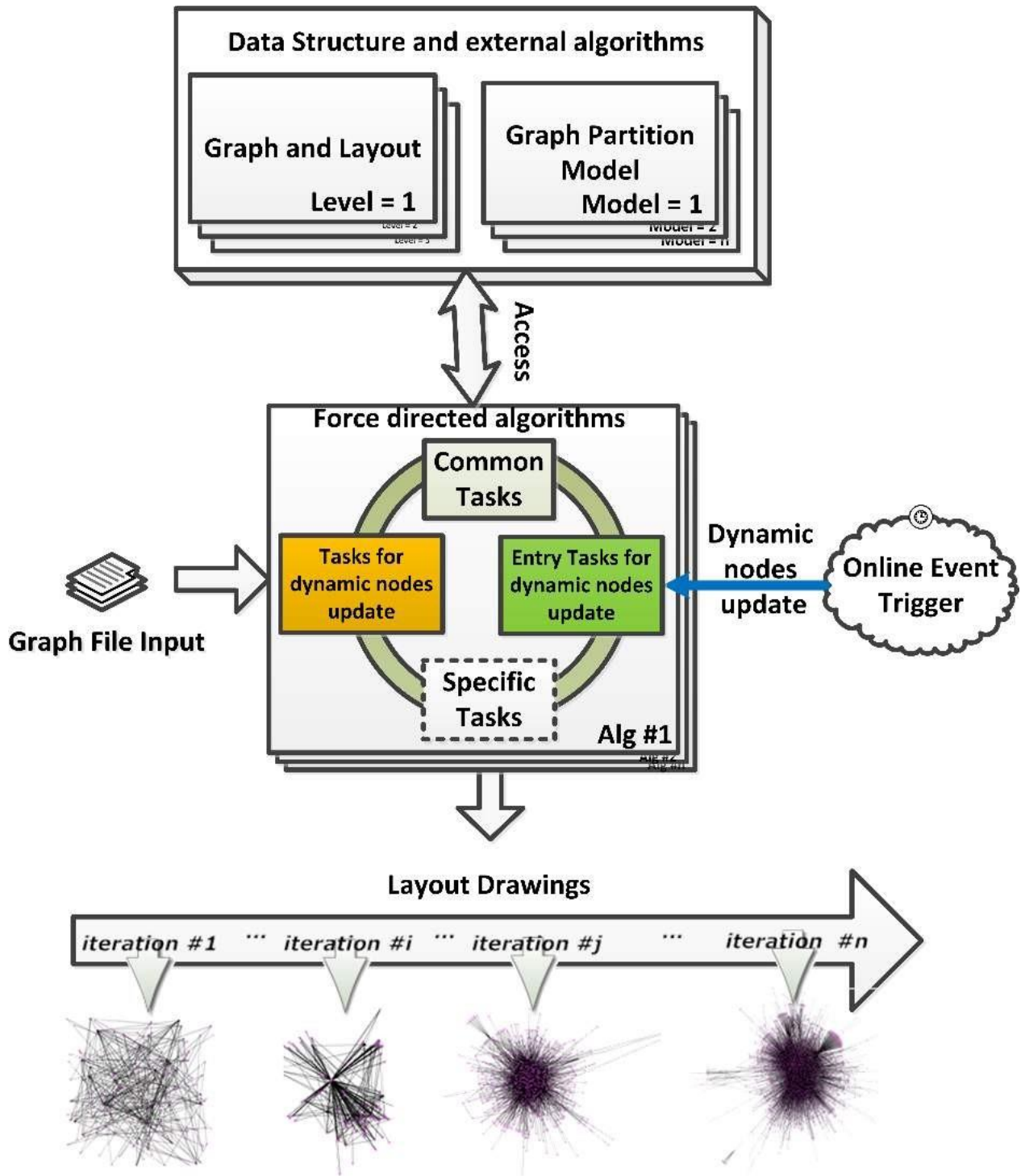

Figure 2: Overview of the design of Online Force-directed Algorithms for visualization of dynamic graphs.

### 3.3.1. Graph File Input

In the proposed architecture, we implement three types of graph inputs: (1) a graph with zero nodes and edges in which the incoming information of nodes and edges are from real-time data steam, (2) a graph with predefined nodes and edges (e.g., from a local graph file) in which the incoming information of nodes and edges are from real-time data steams, and (3) a graph without accepting incoming information from real-time data steam, the information of nodes and edges are read from local graph files.

*3.3.2. Data structures*

Several attributes of nodes and edges can be used in the graph visualization. For example, a node may be associated with the following attributes: associated edges, neighboring nodes, the position of node, whether the position of node is fixed, the label of node, the shape of node, the icon of node, etc. An edge can associated with the following attributes: source node, target node, directed/undirected, whether it is a curve edge, label, weight, etc. A large number of attributes and functions can be considered for the visualization of dynamic graphs. In the proposed design of OFD algorithms, associated edges of nodes, position of nodes, source and target nodes of edges are considered as basic attributes. Other attributes are not implemented in the OFD algorithms. Moreover, we also implement the Java interface [35] of nodes, edges, graphs and layout. Additional attributes can be supported by implementing these interfaces. In this paper, we aim to minimize the effort required to implement these interfaces without committing to details such as whether the graph is a directed graph.

*3.3.3. Force-directed Algorithms*

The core idea of FD algorithms (both SFD and OFD algorithms) is to produce a "good" layout in order to meet the objective of visualization of dynamic graphs. The definition of good layout can be variant depending on the requirement and objective of visualizations. For example, it can be edge-crossing minimization, producing planer graphs, boundary node detection, clustering classification, etc. Therefore, the determination of a good layout splits the control flow of FD algorithms into two branches. If a good layout has been produced, the FD algorithms will be terminated to indicate the objective of visualization has been achieved. Otherwise, an iterative process is used to compute forces and update the position of nodes of the graph. A simplified control flow of force-directed algorithms is illustrated in Figure 3.

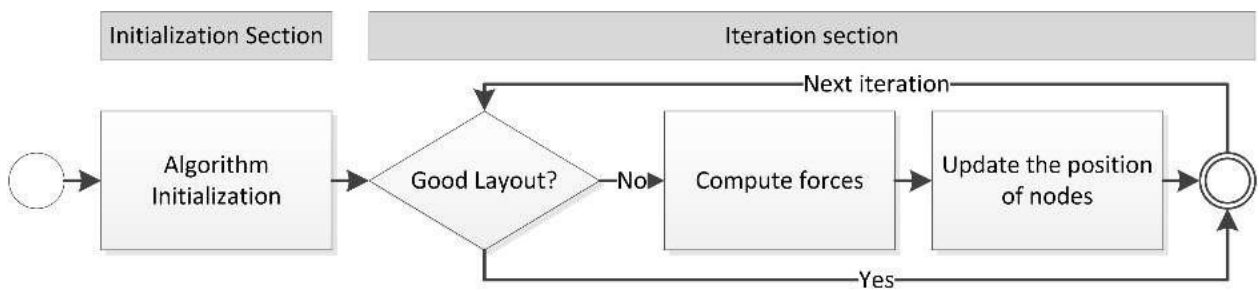

Figure 3: A simplified control flow of force-directed algorithms.

*3.3.4. Online Event Trigger*

Two online event triggers are implemented in the proposed architecture to simulate the tick of the clock, data arrival rate, time constraints per iteration, etc. They are the internal clock ($S_1$) and the WebSocket interface ($S_2$). The internal clock $S_1$ is useful for the simulation of data arrival rate. It uses a timer to update dynamic graphs. When the $S_1$ exceeds a retention period, the online event will be triggered to insert nodes and edges to the dynamic graphs. The WebSocket interface ($S_2$) is used for listening real-time data streams.

For example, real-time transactions from blockchain [2] or from stock trading are the example data streams of WebSocket interface. Every time a packet is received from the WebSocket interface, the FD algorithm is interrupted. Nodes and edges from the received packet are then inserted to the dynamic graph.

*3.3.5. Layout Drawings*

Two types of layout output could be used for the visualization of dynamic graphs: snapshot layout and result layout. The snapshot layout is the intermediate layout from the OFD algorithms. The result layout is the final layout produced at the time the algorithm has been terminated. An example layout of online FR algorithm is illustrated in Figure 4. This visualization is generated from a dataset of a transaction graph from BitCoin Alpha[36].

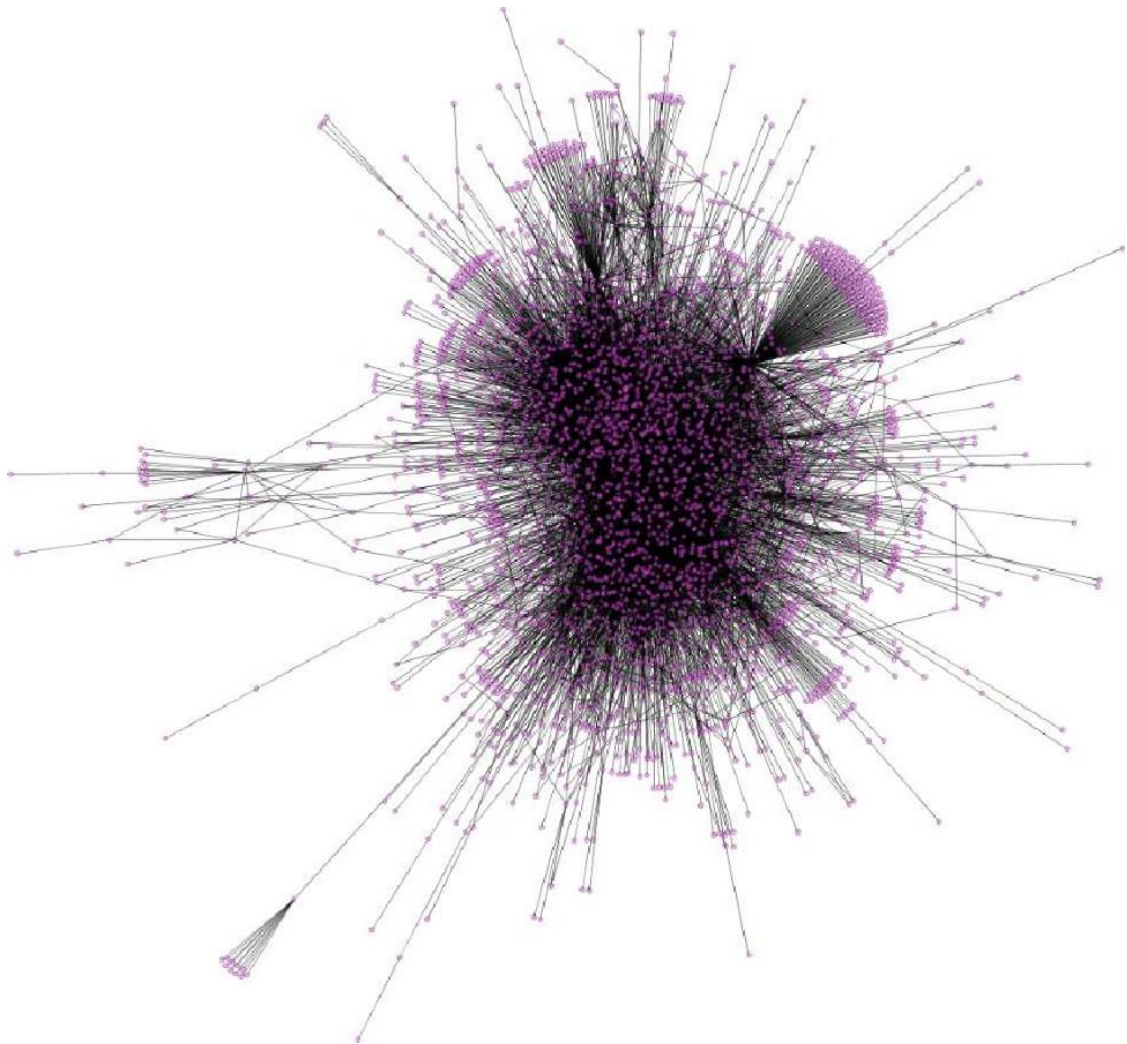

Figure 4: An example layout of a transaction graph from BitCoin Alpha generated by online FR algorithm.

---

[2]https://blockchain.info/api/api websocket

### 3.3.6. Placement of Nodes

In this architecture, we have implemented two approaches to assign the position of nodes. In the first approach, the positions of existing nodes remain unchanged and the positions of new nodes are assigned in a random manner as shown in Figure 5. In this approach, we assume that the position of new nodes are unknown and therefore, random positions are assigned to the new nodes.

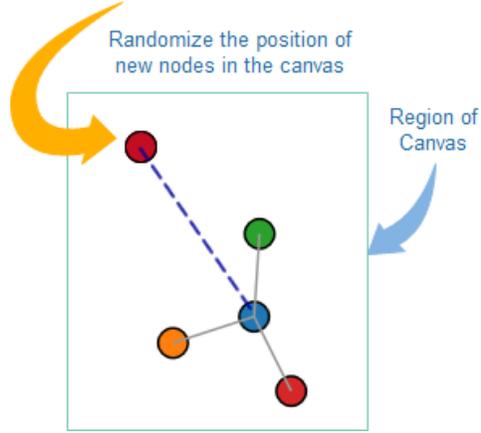

Figure 5: Randomization of nodes.

In the second approach, a node is placed according to the barycenter of its surrounding neighbors. Barycenter is the point of two or more nodes that are orbiting each other. If a node has been placed on the barycenter of neighbors, it can reach a lower energy of the node. However, such strategy assumes that a stable layout of the network topology has been calculated whereby the calculation of barycenter rely on the position of neighbor nodes. That means, barycenter approach does not have significant improvement if a large number of new nodes and edges are inserted to the dynamic graphs. The calculation of the distance of a node *i* to the barycenter is defined as follows:

$$r_i = \frac{a}{1 + \frac{m_i}{m_j}} \quad (4)$$

where $r_i$ is the distance from node *i* to the barycenter, *a* is the distance between the center of nodes *i* and *j*. $m_i$ and $m_j$ are the mass of nodes *i* and *j* respectively. The mass of nodes in network topology can be denoted as the importance/size of nodes. Let $PN(u)$ be the neighbors of node *v* in which every node *u* in $PN(u)$ has assigned a position. The position of *v* [16, 37, 38] then can be calculated as follows:

$$x_v = \frac{1}{|PN(u)|} \sum_{u \in PN(u)} x_u \,;\, y_v = \frac{1}{|PN(u)|} \sum_{u \in PN(u)} y_u \quad (5)$$

An example of barycenter approach in a unweighted network topology is illustrated in Figure 6. Nodes *D* is a new node in Figure 6. Nodes *A*, *B* and *C* are already assigned with a position and they are the neighboring nodes of *D*. The x-axis and y-axis positions of node *D* can be calculated by the Equation 5.

### 3.3.7. Canvas Scaling Normalization

Force-directed algorithms select a custom pre-defined canvas size to draw the layout of graphs. However, nodes could be stacked together in large graphs if the canvas size is small. In the proposed architecture, a dynamic scaling of the canvas size is used to alleviative this problem. OFD algorithms are designed to work with varying size of canvases at every iteration. For example, if a large number of new nodes are inserted to the dynamic graph at iteration $t$, the size of canvas at iteration $t$ can be enlarged proportionally to the number of new nodes inserted. Although dynamic scaling can alleviative nodes stacked together in the canvas, it cased the mismatch of scaling in the animation of multiple layouts because every iteration could have a different canvas size. An example is shown in Figure 7. Suppose the size of canvas at iteration $i$, $j$ and $k$ are $5 \times 5$, $20 \times 5$ and $2 \times 2$ respectively. The layout present in the animation will shift or magnify because the size of canvas at every iteration is different. Therefore, we perform the normalization of scaling of layouts in which a layout with the maximum number of nodes and edges are selected as the "standard layout". For those layouts which appear prior or later than the "standard layout", nodes are shifted or magnified in order to match the identical scaling of the "standard layout".

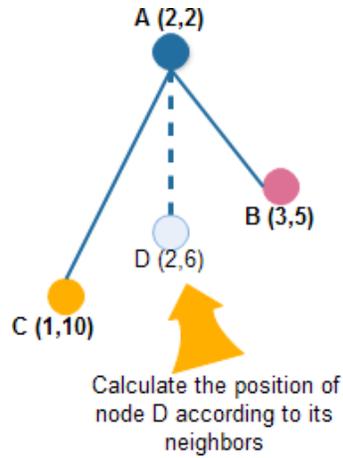

Figure 6: Position calculation by barycenter of neighbor nodes.

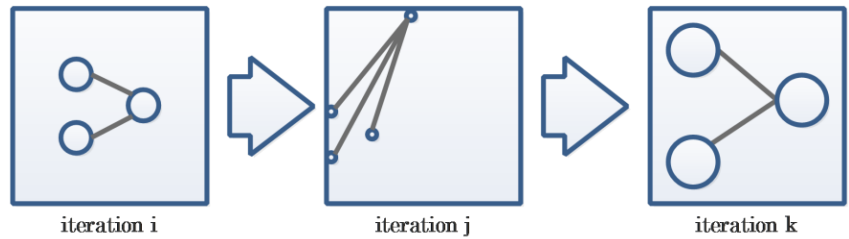

Figure 7: Different scaling of layout in different iterations.

*3.4. Example: Online Kamada-Kawai (KK) Algorithm*

The section describe the detailed design of an OFD algorithm by reusing the force model of the original SFD algorithm. Due to the page limit, we select KK algorithm as an example. The Kamada-Kawai (KK) algorithm [3] is a FD algorithm based on Eades' spring-embedder model [39]. The main objective of the KK algorithm is to keep edge crossing to a minimum, and to distribute the nodes and edges uniformly [40]. To achieve these objectives, KK algorithm uses a spring model that minimizes the energy function of the network topology. Nodes are placed by the algorithm such that their distances in the visualization are proportional to their theoretical graphed distances. The energy function of the KK algorithm can be defined as follows:

$$E = \sum_{i=1}^{n-1} \sum_{j=i+1}^{n} \frac{1}{2} k_{i,j} (|p_i - p_j| - l_{i,j})^2 \qquad (6)$$

where $p_i$ and $p_j$ are the positions of nodes *i* and *j* in the visualization. $k_{i,j}$ is the stiffness of the spring between nodes *i* and *j*. KK algorithm calculates a position for each pair of nodes, *i* and *j*, in the visualization; the Euclidean distance of the pair is proportional to $l_{i,j}$.

Three types of graph inputs are implemented in the proposed architecture. Moreover, a graph could be partitioned into multiple level/sub graphs in SFD algorithms (e.g., LinLog algorithm). Therefore, the input of OFD algorithms contains the graph *G* and the level (*Level*) of sub graphs. In our experiments, we do not count the I/O operations of accessing graph file. In addition, *Level* is set to 1 if the SFD algorithms do not support multiple level/sub graphs.

Force-directed algorithms can be generalized into initialization and iteration sections as we described in Section 3.2. Likewise, the main components of KK algorithm can be summarized as: (1) inputs, (2) the memory allocation for internal data structures, (3) the content of initialization section, (4) the content of iteration section, and (5) output. The pseudo code for the content of KK algorithm is given in Algorithm 1. Moreover, the description and corresponding algorithms for initialization and iteration section of KK algorithm are described in Section 3.4.1 and 3.4.2 respectively.

---

**Algorithm 1:** Pseudo code of Kamada-Kawai algorithm

**Input :** Graph *G*, Level of sub graph *Level*

**Output:** Layout drawings of graph *G*

// 2) Allocate memory for internal data structures

1 *define theoretic graphed distance $d_{i,j} = \varphi$ ;*

2 *define ideal spring distance $l_{i,j} = \varphi$ ;*

3 *define stiffness $k_{i,j} = \varphi$ ;*

// 3) Initializatin Section

4 $G \leftarrow$ init(*G, $d_{i,j}$, $l_{i,j}$, $k_{i,j}$*) *by Algorithm 2*

// 4) Iteration Section

5 $G \leftarrow$ iteration(*G, $d_{i,j}$, $l_{i,j}$, $k_{i,j}$*) *by Algorithm 4*

// 5) Output

6 **return** the layout drawings of graph *G*;

---

*3.4.1. Initialization Section of Online KK Algorithm*

The KK algorithm initializes the ideal spring distance $l_{i,j}$, theoretic graphed distance $d_{i,j}$, stiffness $k_{i,j}$, and the initial position of nodes. The pseudo code for initialization section of KK algorithm is given in

Algorithm 2. Moreover, the definition of ideal spring distance $l_{i,j}$ between nodes $i$ and $j$ is defined in the equation 7 and the corresponding pseudo code is given in Algorithm 3.

$$l_{i,j} = \frac{L_0}{\max_{i<j} d_{i,j}} \times d_{i,j} \qquad (7)$$

---

**Algorithm 2:** Pseudo code for initialization section *init*() of KK algorithm

**Input :** Graph $G$, ideal spring distance $l_{i,j}$, theoretic graphed distance $d_{i,j}$, stiffness $k_{i,j}$

**Output:** Graph $G$

1 **foreach** $i \in V$ **do**
2    // 1) Ideal spring distance
   $l_{i,j} \leftarrow$ *InitLij*($G, i$) by Algorithm 3 ;
3    **foreach** $j \in V$ **and** $j \mathrel{/}= i$ **do**
     // 2) Theoretic graphed distance
4      $d_{i,j} \leftarrow$ hop counts between nodes $i$ and $j$ ;
     // 3) Stiffness
5      $k_{i,j} \leftarrow$ stiffness between nodes $i$ and $j$, default is 1 ;
6    **end**
   // 4) Initial position of nodes
7    initialize position for node $i$ ;
8 **end**
9 **return** $G$;

---

**Algorithm 3:** Pseudo code for ideal spring distance initialization InitLij()

**Input :** Graph $G$, node $i$

**Output:** a matrix of idea spring distance $l_{i,j}$

1 compute the diameter $L$ of $G$ ;
2 let $H$ and $W$ be the height and width of the canvas ;
3 $L_0 \leftarrow \min(H, W)$ ;
4 $L \leftarrow \frac{L_0}{L}$ ;
5 **foreach** $j \in V$ **and** $i \mathrel{/}= j$ **do**
6    let $d_{ij}$ and $d_{ji}$ be the hop count between nodes $i$ and $j$ ;
7    $l_{i,j} \leftarrow \min(d_{ij}, L)$ ;
8    $l_{j,i} \leftarrow \min(d_{ji}, L)$ ;
9 **end**
10 **return** $l_{i,j}$;

*3.4.2. Iteration Section of Online KK Algorithm*

In the iteration section, KK algorithm determines a visual position for every node *v* in the graph *G* and tries to decrease the energy function in the whole network. The pseudo code for the iteration section is given in Algorithm 4.

---
**Algorithm 4:** Pseudo code for iteration section *iteration*() of KK algorithm
---
**Input :** Graph *G*, theoretic graphed distance $d_{i,j}$, ideal spring distance $l_{i,j}$, stiffness $k_{i,j}$
**Output:** Graph *G*

1 **while** $max_i \Delta_i > \epsilon$ **do**
2     let $node_m$ be the node satisfying $\Delta_m = max_i CalcDeltaM(i)$ by Algorithm 5 ;
3     **while** $\Delta_m > \epsilon$ **do**
4        compute $\delta_x$ and $\delta_y$ for $node_m$ ;
        // Seek a visual position for every node $node_m$
5        $x_m \leftarrow x_m + \delta_x$ ;
6        $y_m \leftarrow y_m + \delta_y$ ;
7     **end**
8 **end**
9 **return** *G*;

---

However, solving all of these non-linear equations simultaneously is infeasible because they are dependent on one another. Therefore, an iterative approach can be used to solve the equation based on the Newton-Raphson method. At each iteration, KK algorithm chooses a node *m* which has the largest energy. In other words, the node *m* is moved to the new position, where it can reach a lower level of $\Delta_m$ than prior. Meanwhile, the other nodes remain fixed. The $\Delta_m$ calculation is defined in the equation 8 and its pseudo code is given in Algorithm 5.

$$\Delta m = \sqrt{\left(\frac{\partial E}{\partial x_m}\right)^2 + \left(\frac{\partial E}{\partial y_m}\right)^2} \tag{8}$$

*3.4.3. Specific Tasks of Online KK Algorithm*

In the proposed design, OFD algorithms are inherited from existing SFD algorithms. There is a specific task to be implemented in SFD algorithms to handle dynamic graphs. This task is responsible for the entry point when dynamic nodes and edges are updated. We outline the the pseudo code of the specific task of online KK algorithm in Algorithm 6.

Online KK algorithm is interrupted at the time when the dynamic graph was updated. The specific tasks of online KK algorithms are then forced to be executed to handle dynamic nodes and edges update. Therefore, the only difference between static and online KK algorithm is the implementation of the specific tasks (highlighted in gray color of the pseudo code of Algorithm 6) of online KK algorithm. The dispatch of interrupt is handled by online event trigger of the proposed design, which is transparent to the implementation of force-directed algorithms. Other tasks of online KK algorithm are identical to static KK algorithm.

**Algorithm 5:** Pseudo code for $\Delta_m$ calculation CalcDeltaM()

    **Input** : a node $m$ in graph $G$

    **Output:** the $\Delta_m$ for node $m$

1   let $L$ be the diameter of the graph $G$ ;

2   let $K$ be the stiffness constance, the default value is 1. ;

3 **foreach** $i \in V$ **and** $i \;/\!= m$ **do**

4     $l_{i,m} \leftarrow L \times l_{i,m}$ ;

5     $k_{i,m} \leftarrow \frac{K}{l_{i,m}^2}$ ;

6     $\Delta_x \leftarrow x_m - x_i$ ;

7     $\Delta_y \leftarrow y_m - y_i$ ;

8     $d \leftarrow \sqrt{\Delta_x^2 + \Delta_y^2}$ ;

9     $\Delta x_m \leftarrow \Delta_x \times k_{i,m} \times (1 - \frac{l_{i,m}}{d})$;

10    $\Delta y_m \leftarrow \Delta_y \times k_{i,m} \times (1 - \frac{l_{i,m}}{d})$;

11 **end**

12 **return** $\sqrt{\Delta x_m^2 + \Delta y_m^2}$;

**Algorithm 6:** Pseudo code of online Kamada-Kawai algorithm

    **Input** : Graph *G*, Level *Level*

    **Output:** Layout drawings of graph *G*

    // 2) Allocate memory for internal data structures

1   *define theoretic graphed distance $d_{i,j} = \varphi$* ;

2   *define ideal spring distance $l_{i,j} = \varphi$* ;

3   *define stiffness $k_{i,j} = \varphi$* ;

    // 3) Initializatin Section

4   *$G \leftarrow init(G, d_{i,j}, l_{i,j}, k_{i,j})$ by Algorithm 2*

5   // 4) The entry point when nodes and edges are inserted to the dynamic graph

6   *let $V'$ be the new nodes to be added in G* ;

7   *let $E'$ be the new edges to be added in G* ;

8   **foreach** $j \in V$ **do**

9       **foreach** $x \in V'$ **do**

10         *$l_{x,j} \leftarrow InitLij(G, j)$ by Algorithm 3* ;

11         *$d_{x,j} \leftarrow$ hop counts between nodes x and j* ;

12         *$k_{x,j} \leftarrow$ stiffness between nodes x and j* ;

13         *initialize position for node x* ;

14     **end**

15 **end**

    // 5) Iteration Section

16   *$G \leftarrow iteration(G, d_{i,j}, l_{i,j}, k_{i,j})$ by Algorithm 4*

    // 6) Output

17 **return** the layout drawings of graph *G*;

## 4. Experiments

In this section, we evaluate the performance between SFD and OFD algorithms for visualization of dynamic graphs. We selected five SFD and OFD algorithms for evaluation. Each group of algorithms are based on FR, KK and FA2, DH and LinLog algorithms. All algorithms are implemented in Java programming language. In our experiments, to achieve fair comparison, we do not perform additional optimizations for SFD and OFD algorithms. Specifically, we do not apply any heuristics and hardware accelerating approaches for SFD and OFD algorithms. In addition, SFD and OFD algorithms were executed in sequential processes to eliminate the performance impact caused by multi-threading or concurrent executions. Therefore, all SFD and OFD algorithms are executed on one thread. To satisfy these requirements, we have modified the implementation of SFD and OFD algorithms involving multi-threading into single-threaded procedures.

*4.1. Experiment Settings*

Experiments were performed on a computer with an Intel Xeon E5-2683 v3 processor, 128 GB of memory and FreeBSD 11.0. We used four criteria for the performance evaluation of experiments: 1) execution time, 2) edge crossing, 3) variance of edge crossing and 4) standard deviation of edge lengths. The descriptions of criteria are shown in Table 1. The experimental result of execution time and edge crossing is summarized in Section 4.2. The experimental result of execution time and standard deviation of edge lengths is summarized in Section 4.3.

Table 1: Performance evaluation criteria.

| Criteria | Description | Evaluation |
| --- | --- | --- |
| Execution Time ($T$) | Total amount of execution time that the algorithm ran in seconds | The lower the better |
| Edge Crossing ($EC$) | Total number of pairs of edges which intersect | The lower the better |
| Variance of Edge Crossing ($EC_{VM}$) | The measurement of difference between the high and low values of edge crossing | The lower the better |
| Standard Deviation of Edge Lengths ($EC_{SD}$) | The measurement of variation of edge lengths | The lower the better |

In addition, we selected three datasets with varying size, temporal edges and duration of time interval in our experiments. The detailed information of datasets is shown in Table 2. Dataset email-Eu-core-temporal-Dept3 ($D1$) [41] is an incoming and outgoing traffic for E-Mail graph. A separate edge is created for each recipient of the e-mail. Dataset CollegeMsg ($D2$) [42] is a temporal graph comprised of private messages sent on an online social network at the University of California. Dataset soc-sign-bitcoinalpha ($D3$) [36] is a transaction graph in BitCoin Alpha. Dataset $D1$ [41] has smallest number of nodes among the datasets. Dataset $D2$ [42] has the highest updates and large number of edges. Dataset $D3$ [36] have the largest number of nodes and edges but with small updates.

In our experiment, each $SFD$ and $OFD$ algorithm was executed for 300 seconds. Moreover, we used three distribution functions (e.g., Gaussian distribution, Poisson distribution and random distribution) to simulate the time intervals for updating (insertion of new nodes and edges) the graphs. The relationship among the updates and the insertion of nodes and edges are illustrated in Figure 8. The time interval $t_1$, $t_2$,..., $t_n$ are generated by distribution functions where $n$ is the number of new nodes to be added which is defined in the dataset. In Figure 8, blue color nodes and edges are inserted to the graph after time $t_i$ has elapsed. For the case of dataset $D3$ [36], we simulate 1600 updates based on the approach illustrated in Figure 8. The updates for the remaining datasets from the experiments are simulated in a similar way. Moreover, the timestamp and the total duration in each dataset are different. Because of this, we normalize of the duration of datasets based on Equation 9.

Table 2: Properties of datasets.

| Abbr. | Dataset Name | Nodes | Edges | Updates |
|---|---|---|---|---|
| $D1$ | email-Eu-core-temporal-Dept3 [41] | 89 | 12216 | 8822 |
| $D2$ | CollegeMsg [42] | 1899 | 59835 | 58861 |
| $D3$ | soc-sign-bitcoinalpha [36] | 3783 | 24186 | 1600 |

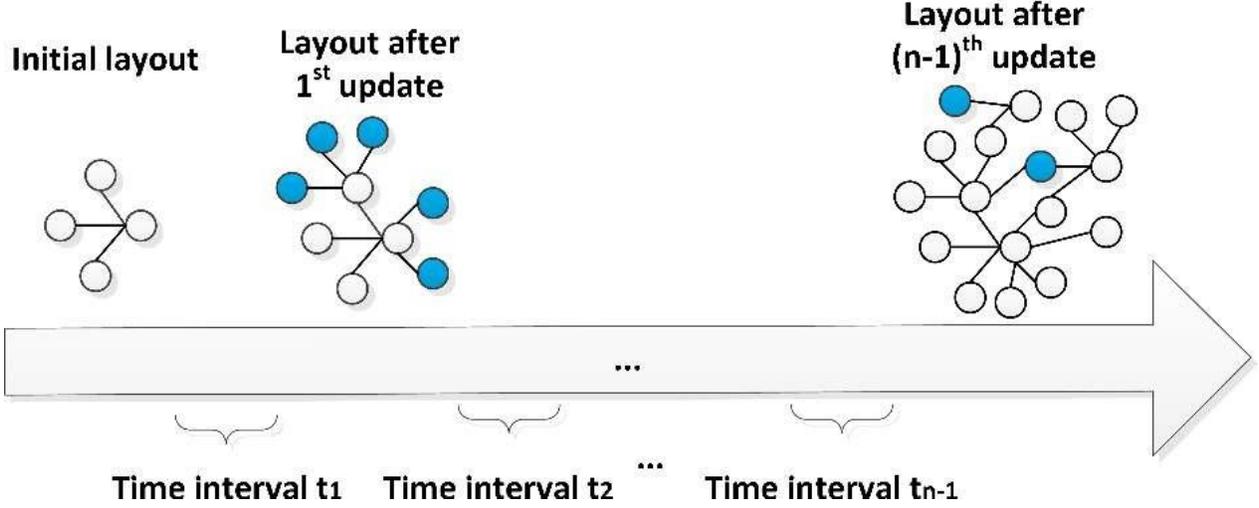

Figure 8: Relationship among updates and the insertion of nodes and edges.

$$z_i = \frac{x_i - min(x)}{max(x) - min(x)} \times 300 \qquad (9)$$

where $x = (x_1, ...x_n)$ are the timestamps of the dataset. $min(x)$ and $max(x)$ are the minimum and maximum timestamp of the dataset. $z_i$ is the normalized timestamp between 0 and 300 seconds.

We defined the variance of edge crossing ($EC_{VM}$) in Equation 10.

$$EC_{VM}(f) = max(EC_f) - min(EC_f) \qquad (10)$$

where $EC_f$ is the number of pairs of intersected edges of algorithm $f$. An example of variance of edge crossing ($EC_{VM}$) is illustrated in Figure 9. Figure 9(a) and Figure 9(b) show a high and low variance of edge crossing observed from the visualizations of experimental results. A low $EC_{VM}$ indicates the steady number of edge crossings during the execution. $EC_{VM}$ is useful for observing whether force-directed algorithms can maintain a stable visualization of dynamic graphs. For example, a high $EC_{VM}$ indicates that the force-directed algorithm cannot preserve a consistent layout of graph when new nodes and edges are inserted to the dynamic graphs.

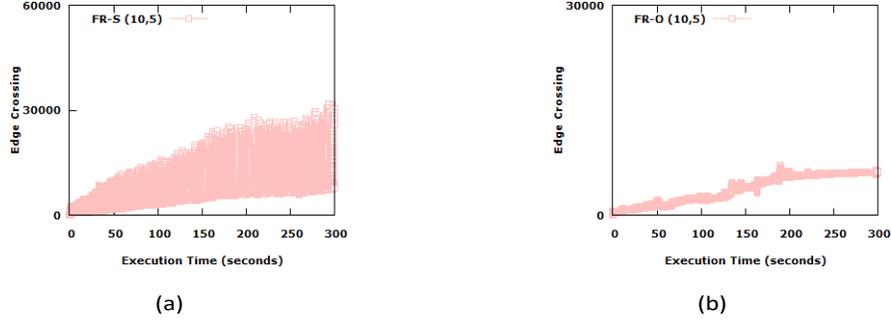

Figure 9: An example of (a) high, (b) low variance of edge crossing.

*Gaussian Distribution Function.* The definition of Gaussian distribution used in our experiments is given in following equation:

$$p(x) = \frac{1}{d1\sqrt{2\pi}} e^{-\frac{1}{2}\left(\frac{x-d}{d1}\right)^2} \qquad (11)$$

where $d$ is the mean and $d1$ is the standard deviation of the Gaussian distribution. We used four sets of parameters of Gaussian distribution to evaluate datasets for our experiments: $d = 10$ and $d1 = 0$, $d = 10$ and $d1 = 1$, $d = 10$ and $d1 = 5$, $d = 20$ and $d1 = 5$. The expectation of the distribution (i.e. mean $d$) starts at 10 and the amount of variation (i.e. standard deviation $d1$) starts from 0 at the first set of parameters. We then keep the mean $d$ unchanged but increase the standard deviation $d1$ at the second and third sets of parameters. Finally, we used a higher value of mean $d$ and standard deviation $d1$ at the last set of parameters of Gaussian distribution.

*Poisson Distribution Function.* The definition of Poisson distribution is shown as follows:

$$p(k) = \frac{u^k}{k!} e^{-d} \qquad (12)$$

where $k$ is a constant and $d$ is the mean of Poisson distribution. We used two sets of parameters of Poisson distribution to evaluate datasets for our experiments. They are $d = 10$ and $d = 20$.

*Random Distribution Function.* We used a uniform deviate [43] which is between range of [*min*, *max*]. Moreover, we used two sets of parameters of random distribution to evaluate datasets for our experiments: 1) *min* = 5 and *max* = 15, and 2) *min* = 15 and *max* = 25.

*4.2. Execution Time vs Edge Crossing*

The experimental results for the execution time versus the number of edge crossings are discussed in this section.

*4.2.1. Edge Crossing Evaluation for Gaussian Distribution*

The experimental results of Gaussian distribution and dataset *D1* [41], *D2* [42] and *D3* [36] are given in Figures 10, 11 and 12 respectively. $FA2 - S$, $FR - S$, $KK - S$, $DH - S$ and $LINLOG - S$ in these figures represent static FA2, **FR,** **KK,** DH and LinLog algorithms. $FA2 - O$, $FR - O$, $KK - O$, $DH - O$ and *LINLOG* $-$ *O* represent online FA2, FR, KK, DH and LinLog algorithms. In each figure, mean *m* and the standard deviation *d* of the Gaussian distribution are shown in parentheses (*m, d*) along with the name of the algorithm evaluated.

From Figure 10, 11 and 12, we can observe that the $EC_{V\ M}$ of OFD algorithms is low for Gaussian distribution meaning that better performance on edge crossings can be achieved by using OFD algorithms. The number of edge crossing of OFD algorithms is almost two times less than SFD algorithms for dataset with small number of nodes. Dataset with high number of nodes and edges, the number of edge crossing for *FA*2 $-$ *O*, *FR* $-$ *O* and *KK* $-$ *O*, *LINLOG* $-$ *O* algorithms are approximately 1 million, 2.5 millions, 0.5 million, 2.4 millions less than the SFD algorithms respectively. However, *KK* $-$ *O* and *DH* $-$ *O* algorithms do not have significant improvement especially when the number of nodes is high. Moreover, *FA*2 $-$ *S* and *LINLOG* $-$ *S* algorithms have better performance in the dataset which has high frequency of updates and number of edges.

*4.2.2. Edge Crossing Evaluation for Poisson Distribution*

The experimental results of Poisson distribution for dataset *D*1 [41], *D*2 [42] and *D*3 [36] are depicted in Figures 13, 14 and 15 respectively. In each figure, mean *m* of Poisson distribution are shown in parentheses (*m*) along with the name of the algorithm evaluated. From the results, we can observe that the $EC_{V\ M}$ of OFD algorithms is low similar to the experimental results shown in Gaussian distribution except *LINLOG* $-$ *O* algorithm. The number of edge crossing for *FR* $-$ *O* and *LINLOG* $-$ *O* algorithm are about 2 and 4 times less than *FR* $-$ *S* and *LINLOG* $-$ *S* algorithms respectively for the dataset with small number of nodes; 3 and 5.4 times less than SFD algorithms respectively for dataset with high updates and number of edges. Moreover, the *FR* $-$ *O* algorithm is about 2 millions edge crossing less than the *FR* $-$ *S* algorithm for dataset with high number of nodes and edges but with small updates. However, *FA*2 $-$ *O*, *KK* $-$ *O* and *DH* $-$ *O* algorithms do not have significant improvements in the Poisson distribution although $EC_{V\,M}$ is lower.

*4.2.3. Edge Crossing Evaluation for Random Distribution*

The experimental results for dataset *D*1 [41], *D*2 [42] and *D*3 [36] with random distribution are depicted in Figures 16, 17 and 18 respectively. In each figure, the ranges of random distribution are shown in parentheses along with the name of the algorithm evaluated.

In this experiment, SFD and OFD algorithms achieve similar number of edge crossings for the dataset with small number of nodes except for the *DH* $-$ *O* algorithm. In the dataset with high number of nodes and edges but with small updates, the number of edge crossing for *FA*2 $-$ *O* and *FR* $-$ *O* algorithms are about 1 million and 2 millions less than their corresponding SFD counterparts. The experimental results also reveal that the $EC_{V\ M}$ of OFD and SFD algorithms are low in the random distribution. Moreover, *KK* $-$ *O* algorithm does not have significant improvements over *KK* $-$ *S* algorithm in all datasets. *DH* $-$ *S* algorithm has better performance in the dataset which has small number of nodes.

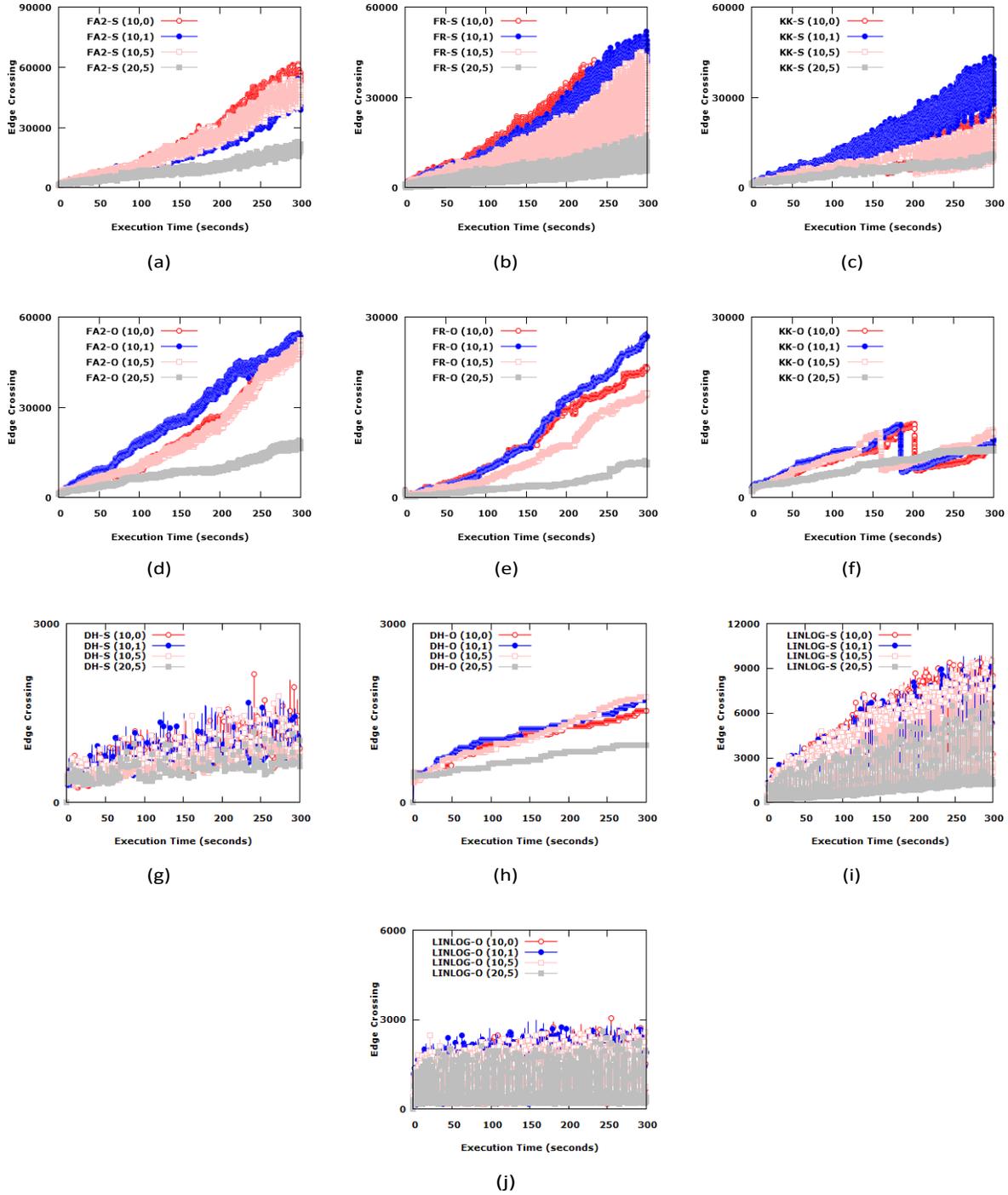

Figure 10: Edge crossing evaluation on dataset $D1$ with Gaussian distribution (a) static FA2, (b) static FR, (c) static KK, (g) static DH, (i) static LinLog (d) online FA2, (e) online FR, (f) online KK, (h) online DH, (j) online LinLog algorithms.

## 4.3. Execution Time vs Standard Deviation of Edge Lengths

In this experiment, we evaluate the performance between SFD and OFD algorithms based on the standard deviation of edges length in output visualization. In our evaluation, we plot the standard deviation of edge lengths at every iteration of SFD and OFD algorithms. Specifically, we used the standard deviation of edge lengths as the metrics to measure the distribution of edge lengths in dynamic graphs. Such measurement can tell us if the generated visualization has significant movement of nodes' position over time.

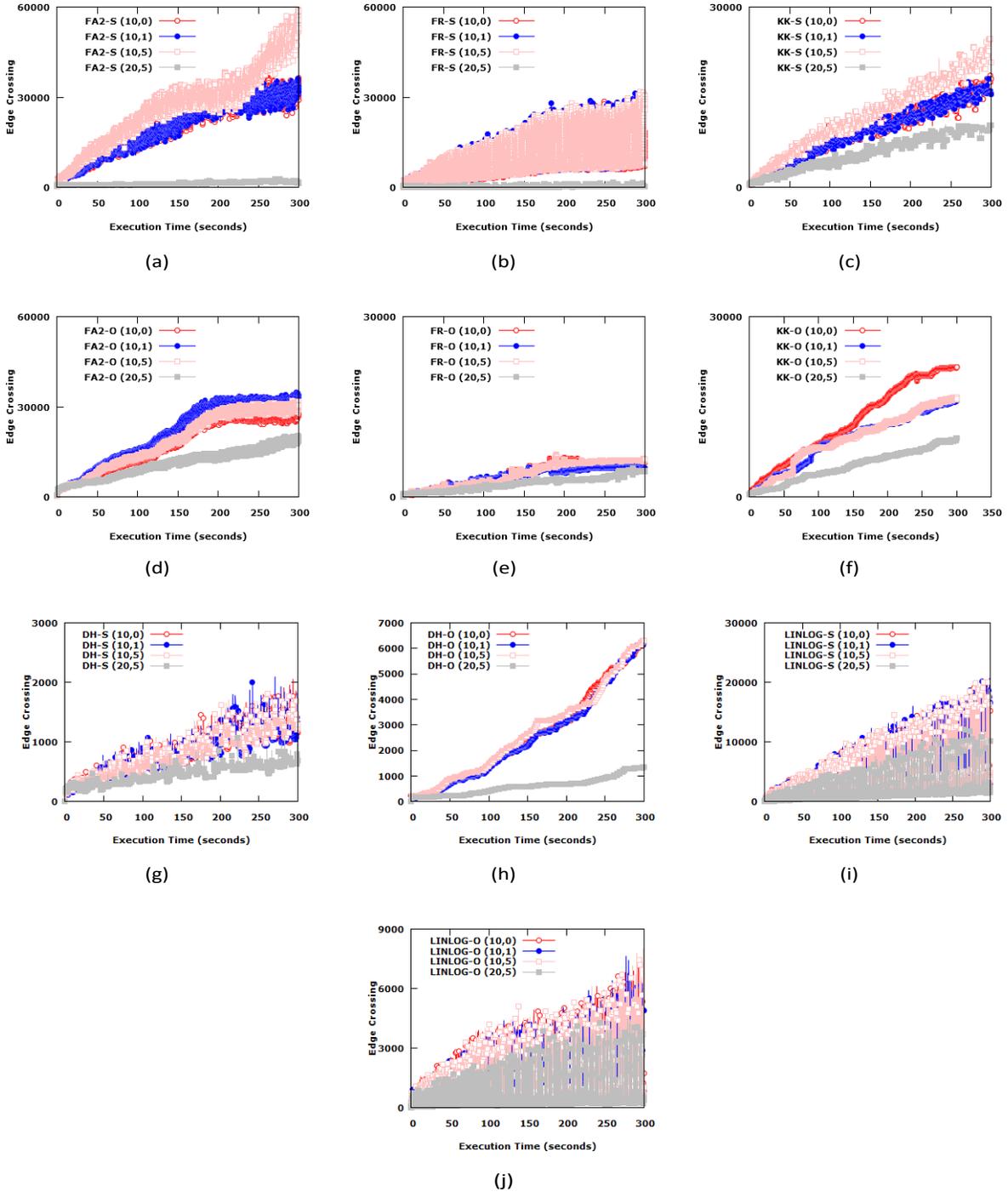

Figure 11: Edge crossing evaluation on dataset $D2$ with Gaussian distribution (a) static FA2, (b) static FR, (c) static KK, (g) static DH, (i) static LinLog (d) online FA2, (e) online FR, (f) online KK, (h) online DH, (j) online LinLog algorithms.

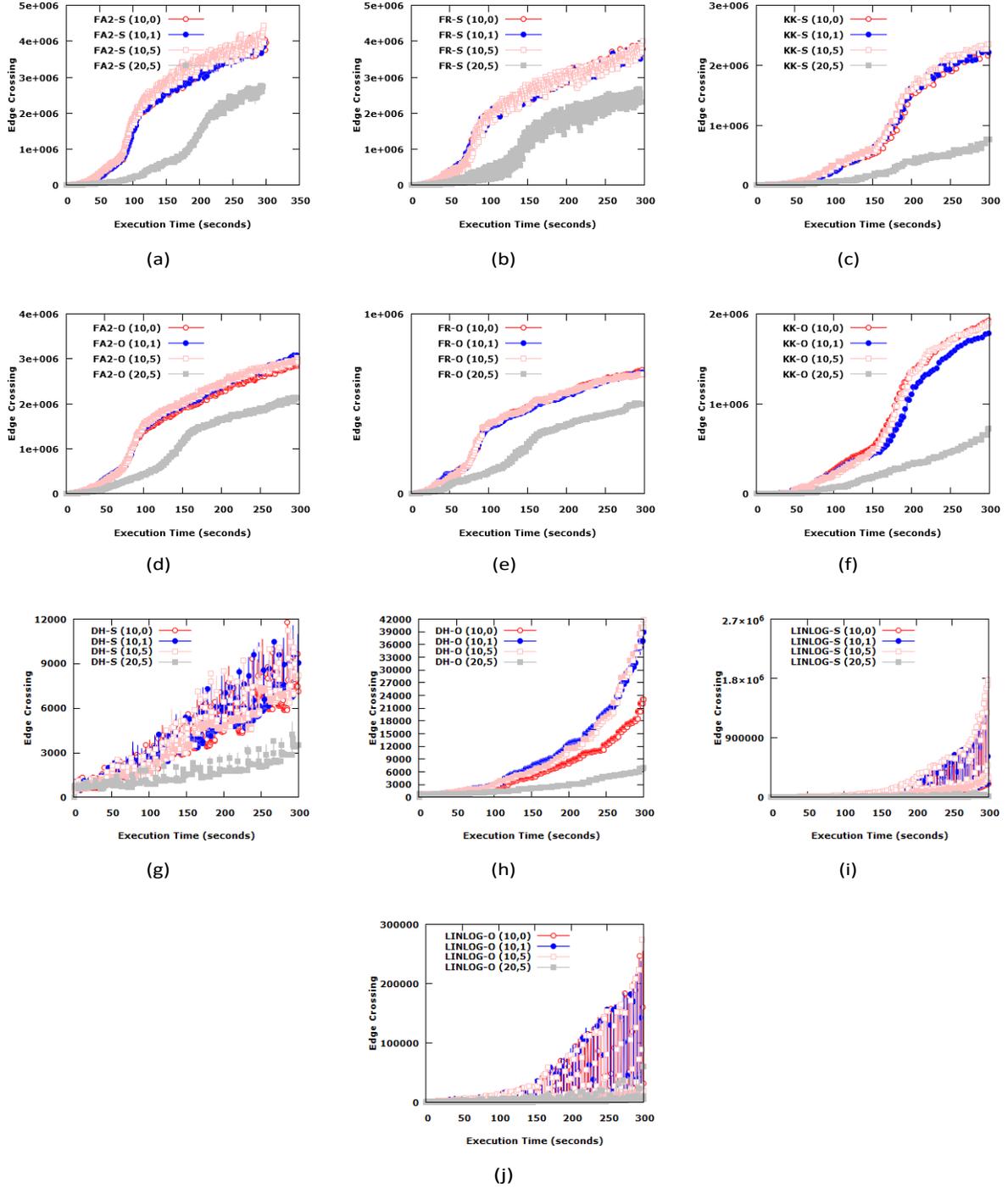

Figure 12: Edge crossing evaluation on dataset $D3$ with Gaussian distribution (a) static FA2, (b) static FR, (c) static KK, (g) static DH, (i) static LinLog (d) online FA2, (e) online FR, (f) online KK, (h) online DH, (j) online LinLog algorithms.

### 4.3.1. Standard Deviation of Edge Lengths Evaluation for Gaussian Distribution

The standard deviation of edge lengths in Gaussian distribution is shown in Figure 19 and Figure 20. According to the experimental results, the standard deviation of edge lengths in $FA2 - S$, $LINLOG - S$, $FA2 - O$ and $LINLOG - O$ algorithms are similar. The standard deviation of edge lengths in $DH - O$ algorithm is higher than the $DH-S$ algorithm. However, the results of FR and KK algorithms are rather interesting. Figure 20 shows that the standard deviation of edge lengths in $FR-O$ algorithm is zigzag in the first 100 seconds, which then becomes stable after $230^{th}$ seconds. In $FR-S$ algorithm, the standard deviation of edge length is undulated until 300 seconds meaning that the position of nodes are still updated dramatically at $300^{th}$ seconds. That is, higher the standard deviation the edge lengths have, larger the movement of nodes does. The standard deviation of edge crossing of $KK - O$ algorithm goes up and down sharply for dataset with high number of nodes and edges but with small updates.

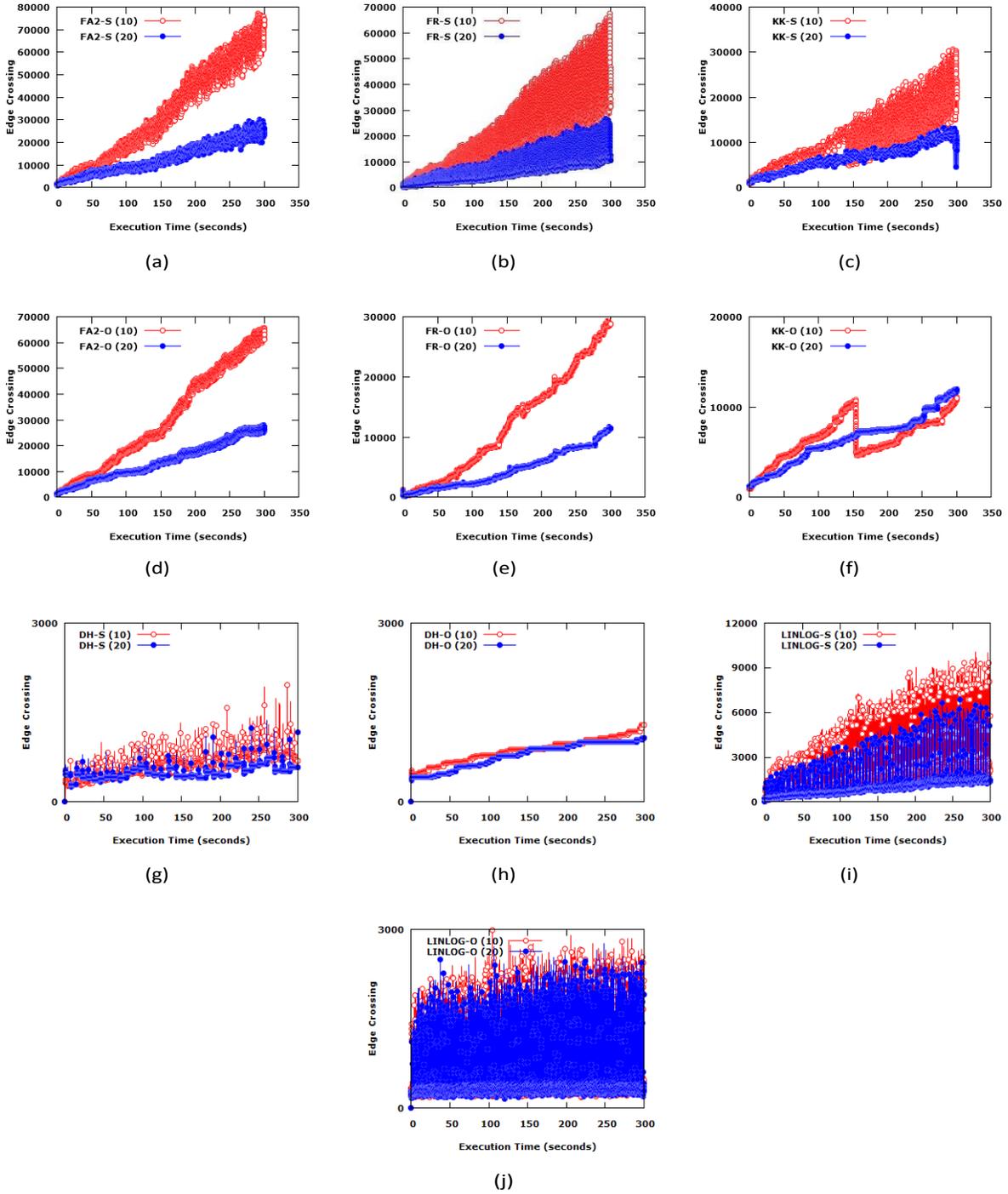

Figure 13: Edge crossing evaluation on dataset $D1$ with Poisson distribution (a) static FA2, (b) static FR, (c) static KK, (g) static DH, (i) static LinLog (d) online FA2, (e) online FR, (f) online KK, (h) online DH, (j) online LinLog algorithms.

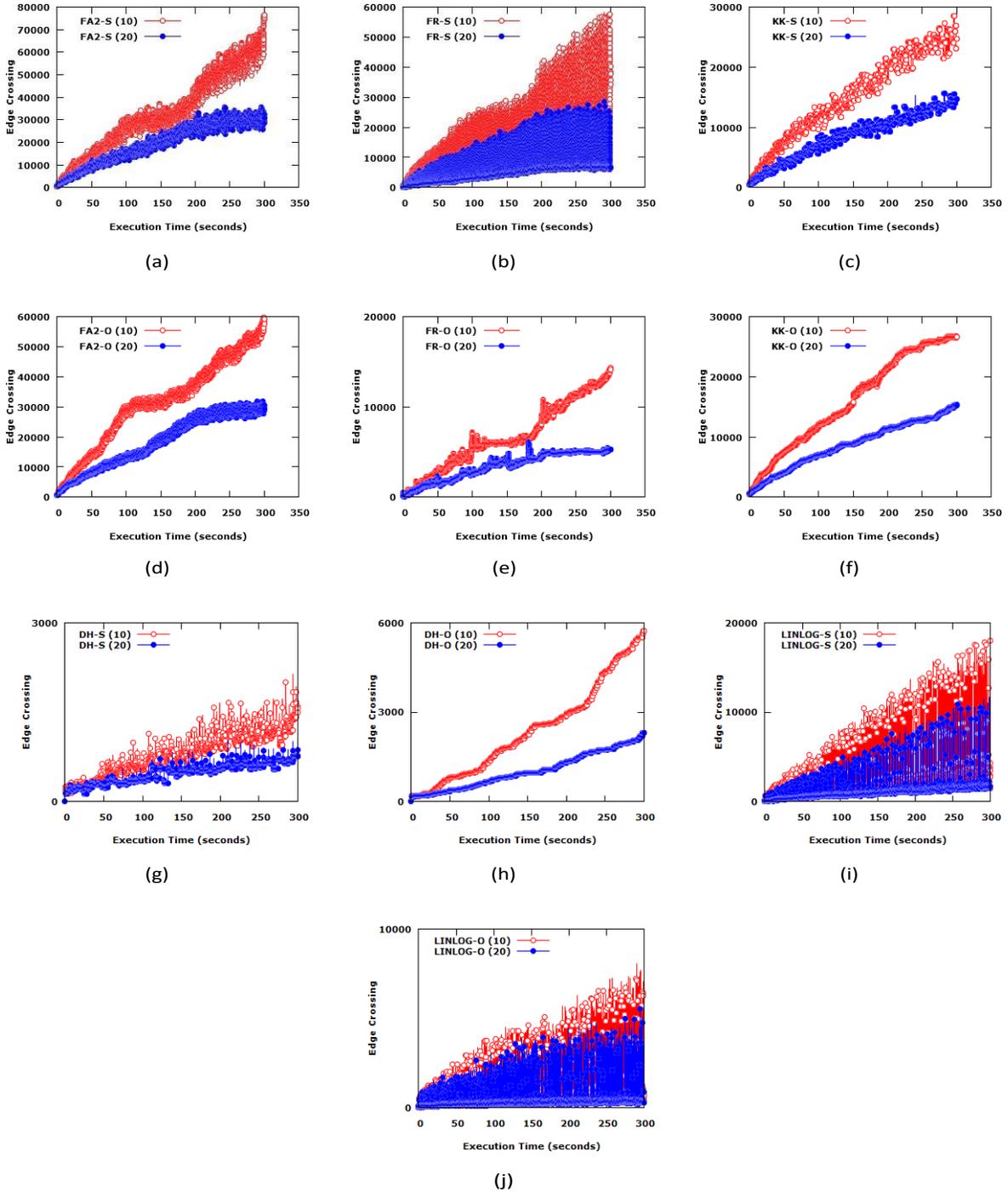

Figure 14: Edge crossing evaluation on dataset $D2$ with Poisson distribution (a) static FA2, (b) static FR, (c) static KK, (g) static DH, (i) static LinLog (d) online FA2, (e) online FR, (f) online KK, (h) online DH, (j) online LinLog algorithms.

In addition, there is a phenomena in the $KK - S$ algorithm in which the standard deviation of edge lengths is always high (no significant drops). Such phenomena does not happen in the $KK-O$ algorithm and other OFD algorithms. Therefore, we extracted and analyzed cases which has high standard deviation of edge lengths from the results of $KK-S$ algorithm. We found that the $KK-S$ algorithm has many cases with high standard deviation of edge lengths especially when the number of nodes becomes large. The standard deviation of edge lengths are high after $150^{th}$ second because a large number of nodes and edges have been inserted into the dynamic graph.

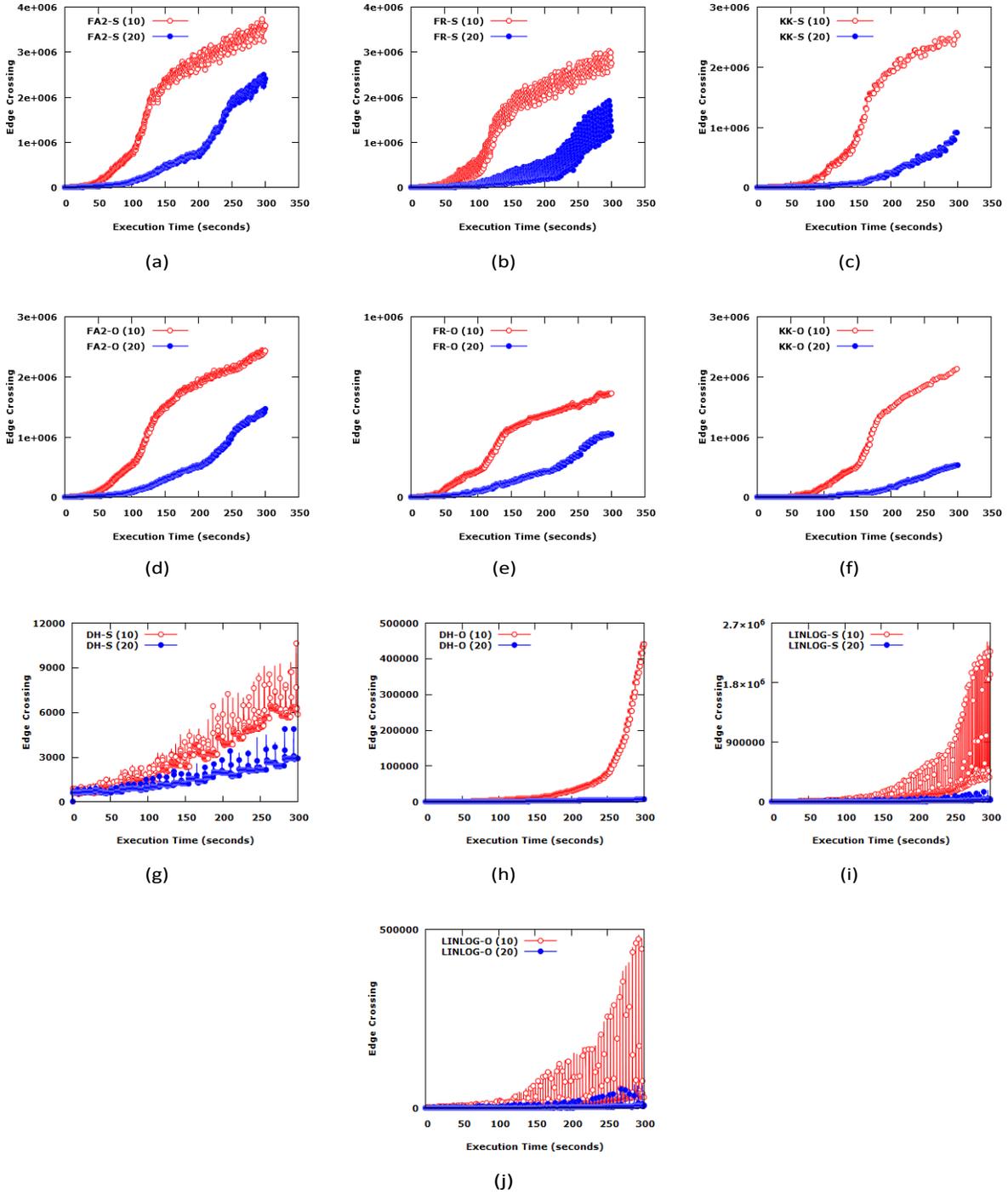

Figure 15: Edge crossing evaluation on dataset $D3$ with Poisson distribution (a) static FA2, (b) static FR, (c) static KK, (g) static DH, (i) static LinLog (d) online FA2, (e) online FR, (f) online KK, (h) online DH, (j) online LinLog algorithms.

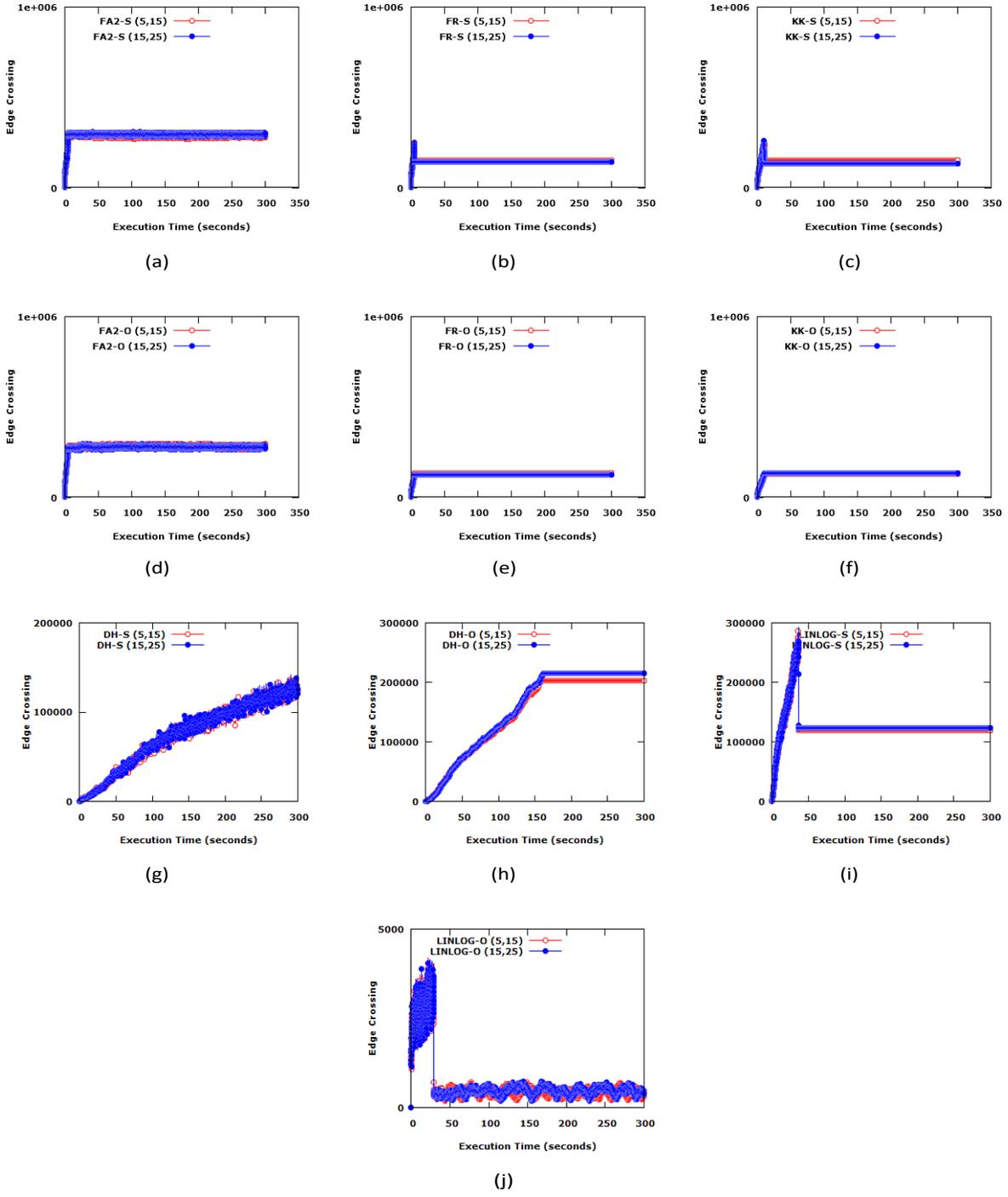

Figure 16: Edge crossing evaluation on dataset $D1$ with random distribution (a) static FA2, (b) static FR, (c) static KK, (g) static DH, (i) static LinLog (d) online FA2, (e) online FR, (f) online KK, (h) online DH, (j) online LinLog algorithms.

We also identified the case which has caused high standard deviation of edge lengths by extracting and analyzing layout drawing from the $KK-S$ algorithm. We found that the $KK-S$ algorithm selects a node in every iteration to update its position in which the selected node could be often pulled too far away. However, $FA2 - S$ and $FR - S$ algorithms update the position of a set of nodes per iteration and hence reduces the chance of having extreme long edge lengths. An example of layout drawing extracted from the $KK - S$ algorithm is illustrated in Figure 21(b). Moreover, high standard deviation of edge lengths often happens at the time when the position of nodes are reinitialized. That is because $KK-S$ algorithm reinitialize the position of nodes in the whole graph at the time new nodes and edges are inserted. In contrast, $KK-O$ algorithm maintains the positions of existing nodes when new nodes and edges are inserted. Therefore, the possibility of high standard deviation of edge lengths are reduced in $KK-O$ algorithm.

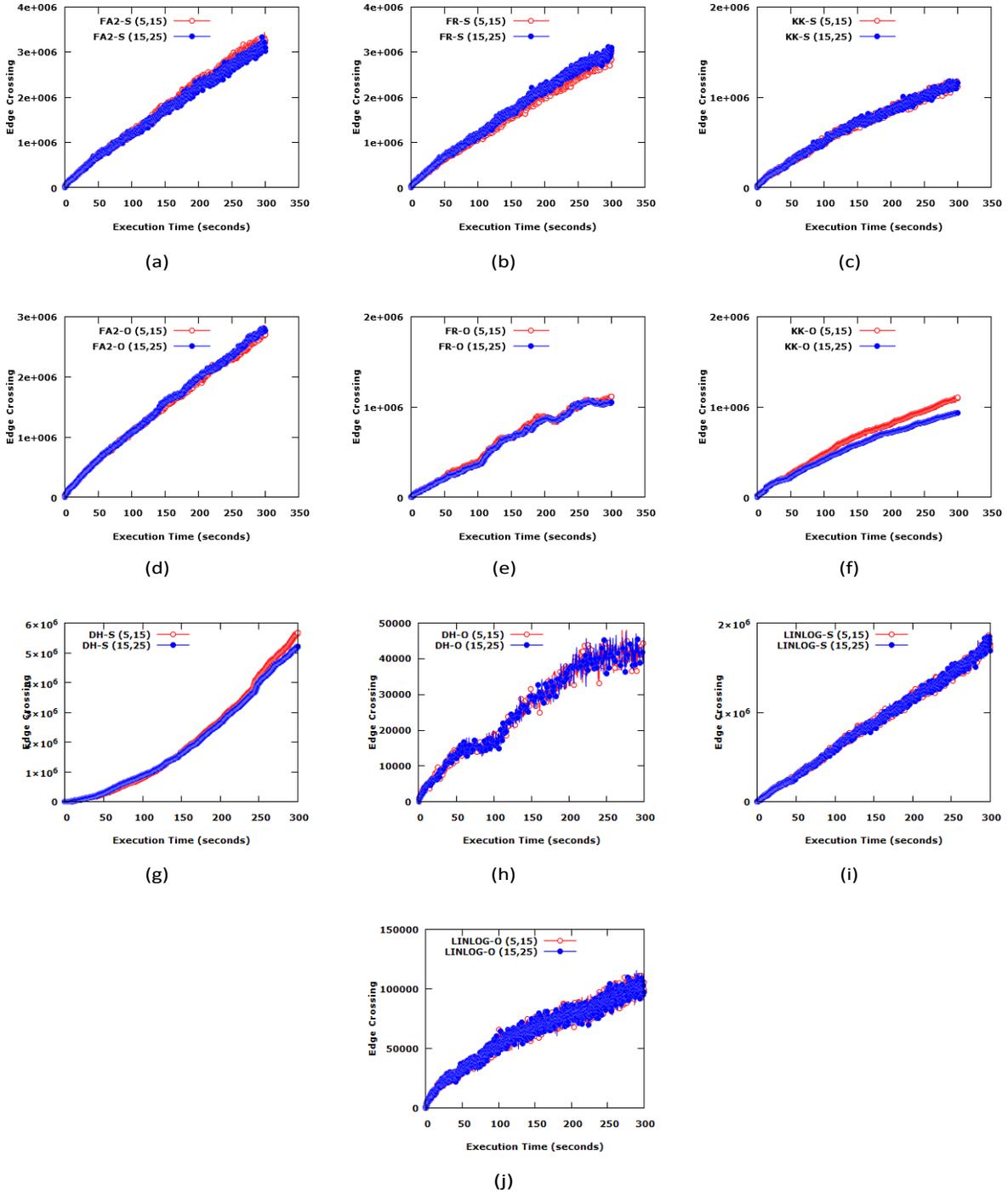

Figure 17: Edge crossing evaluation on dataset $D2$ with random distribution (a) static FA2, (b) static FR, (c) static KK, (g) static DH, (i) static LinLog (d) online FA2, (e) online FR, (f) online KK, (h) online DH, (j) online LinLog algorithms.

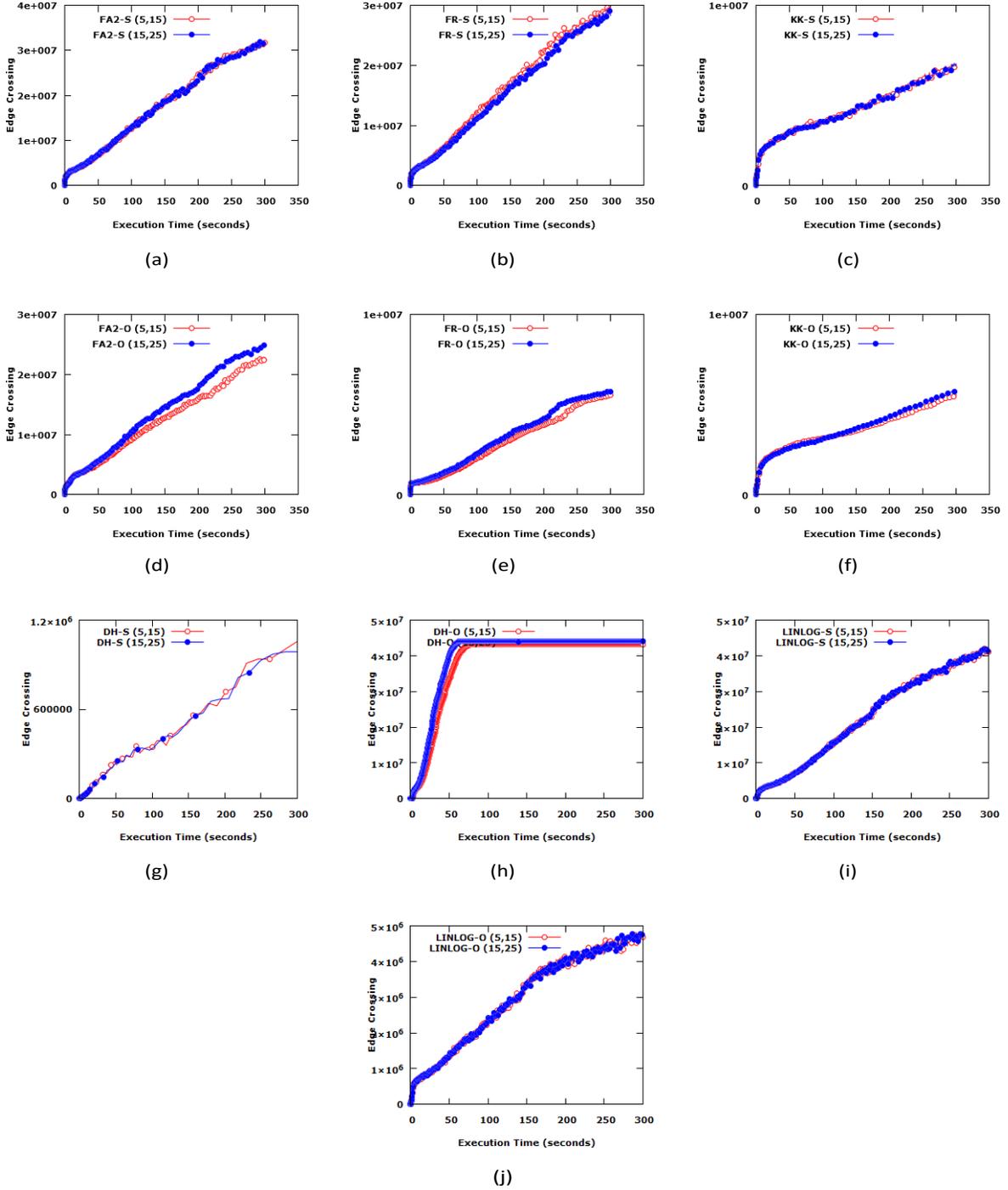

Figure 18: Edge crossing evaluation on dataset $D3$ with random distribution (a) static FA2, (b) static FR, (c) static KK, (g) static DH, (i) static LinLog (d) online FA2, (e) online FR, (f) online KK, (h) online DH, (j) online LinLog algorithms.

*4.3.2. Standard Deviation of Edge Lengths Evaluation for Poisson Distribution*

The standard deviation of edge lengths in Poisson distribution is shown in Figure 22 and Figure 23. The standard deviation of edge lengths in both SFD and OFD algorithms are similar to Gaussian distribution. Specifically, SFD and OFD algorithms have similar standard deviation of edge lengths to Gaussian distribution if the parameter mean *d* is the same in which *d* is used to simulate the interval of updates.

*4.3.3. Standard Deviation of Edge Lengths Evaluation for Random Distribution*

Random distribution can be used to show the performance of SFD and OFD algorithms in the case the interval of updates is long. Because we used a long interval time of update (i.e. to reduce insertion counts) of random distribution in which *min* and *max* are set to 5 and 15, 15 and 25. That means, both SFD and OFD algorithms have longer execution time before new nodes and edges are inserted to the dynamic graph.

The standard deviation of edge lengths in random distribution is shown in Figure 24 and Figure 25. According to the experimental results, the standard deviation of edge lengths in $FR-O$ and $LINLOG-O$ algorithms fluctuate dramatically comparing to the $FR-S$ and $LINLOG-S$ algorithms respectively. In addition, both $KK-O$ and $KK-S$ algorithms have a significant drop of the standard deviation of edge lengths in dataset with small number of nodes. We captured the drop and magnified it in Figure 21(a) for further analysis. According to Figure 21(a), the standard deviation of edge lengths drops at a fast pace in both $KK-S$ and $KK-O$ algorithms because the dataset is a small dynamic graph and the time between each update of node/edge is long enough for both $KK-S$ and $KK-O$ algorithms to produce uniform layout drawings. Moreover, we can observe that $DH-S$ algorithm has significant gaps in Figure 24. That is because a long execution time (7 or more seconds) was needed for the iteration of $DH-S$ algorithm especially in the dataset which has high amount of nodes and edges.

## 5. Conclusion

Visualization of dynamic graphs has been successfully applied in applications such as community and relation exploration in social networks, visualizing metabolic networks, viewing gene structures in bioinformatics, and money market visualization of transaction networks. Although SFD algorithms can be used for visualizing static graphs, they are not suitable for visualization of dynamic graphs. In this paper, we have proposed a novel approach for implementing OFD based on the existing SFD for visualization of dynamic graphs. The proposed approach is based on the results from the comprehensive analysis of common and specific tasks from existing SFD algorithms.

In addition, OFD algorithms allow reusing the force models from SFD algorithms without significant modifications. Therefore, the OFD algorithms developed based on the proposed architecture are able to inherit the properties of SFD algorithms. To the best of our knowledge, our approach is the first systematic and bottom-up design of OFD algorithms in which their force models of can be reused.

Besides, the OFD algorithms are designed with efficient memory management and IO handling strategies for processing a large graph with complex topology. To boost their performances, the algorithms are also designed to reduce the overall re-calculations during the node/edge updating.

In our experiments, we evaluate the performance of SFD and OFD algorithms with datasets of varying sizes. Specifically, we evaluate the performance between SFD and OFD algorithms based on the number of edge crossing, the standard deviation of edge length, variance of edge crossing, and execution time with respect to varying numbers of nodes and average degrees in dynamic graphs.

Experimental results have showed that the OFD algorithms achieve the overall best results in the edge crossing minimization. OFD algorithms have low standard deviation of edge lengths among the datasets. In addition, the number of edge crossing of OFD algorithms is at least two times less than the SFD algorithms for dynamic graphs with a small number of nodes except for the $DH-O$ algorithm. The experiments have also showed that $FR-O$, $FA2-O$ and $LINLOG-O$ algorithms can produce better performance for edge crossing if the data arrival rate (the duration of time interval) of dynamic graphs is high. However, $KK-O$ and $DH-O$ algorithms do not have a significant improvement when the number of nodes is high or the data arrival rate is high.

As for the future work, we are planning to improve the selection of OFD algorithms for the visualization of real-time dynamic graphs. According to the experimental results shown in this paper, the performance of OFD algorithms is found to be related to the characteristics of dynamic graphs. We believe that a more systematic selection of OFD algorithms is needed especially in real-time applications [44].

**Acknowledgement**

This research was funded by the Research Committee of University of Macau, grants MYRG2019-00136-FST and MYRG2017-00029-FST.

[38] S. G. Kobourov, Spring embedders and force directed graph drawing algorithms, arXiv preprint arXiv:1201.3011.

[39] P. Eades, A heuristic for graph drawing, Congressus numerantium 42 (1984) 149–160.

[40] C. Chen, Information visualization: Beyond the horizon, Springer Science & Business Media, 2006.

[41] A. Paranjape, A. R. Benson, J. Leskovec, Motifs in temporal networks, in: Proceedings of the Tenth ACM International Conference on Web Search and Data Mining, WSDM '17, ACM, New York, NY, USA, 2017, pp. 601–610. doi:10.1145/3018661.3018731.
URL http://doi.acm.org/10.1145/3018661.3018731

[42] P. Panzarasa, T. Opsahl, K. M. Carley, Patterns and dynamics of users' behavior and interaction: Network analysis of an online community, Journal of the Association for Information Science and Technology 60 (5) (2009) 911–932.

[43] A. J. Kinderman, J. F. Monahan, Computer generation of random variables using the ratio of uniform deviates, ACM Transactions on Mathematical Software (TOMS) 3 (3) (1977) 257–260.

[44] M. K. Agarwal, K. Ramamritham, M. Bhide, Real time discovery of dense clusters in highly dynamic graphs: identifying real world events in highly dynamic environments, Proceedings of the VLDB Endowment 5 (10) (2012) 980–991.
38

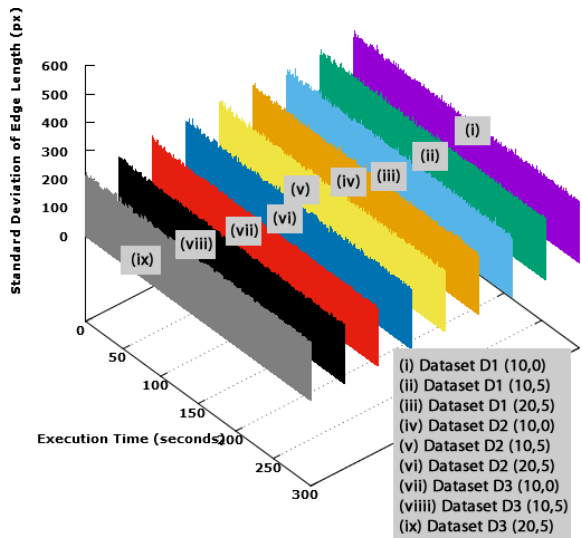
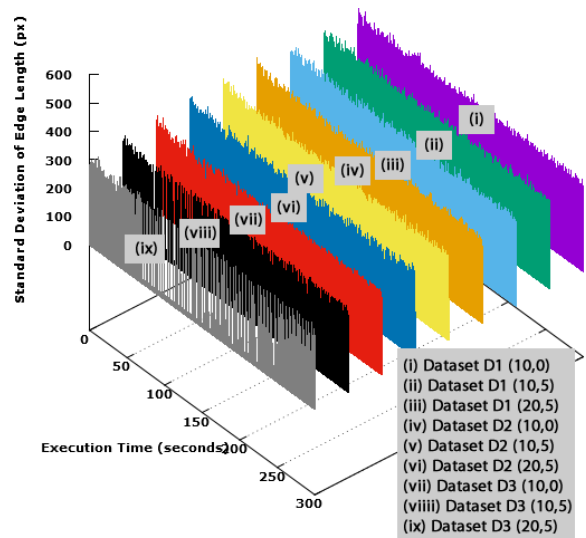
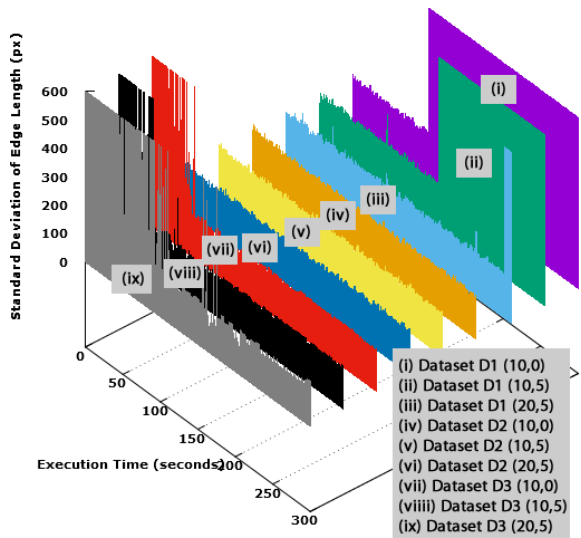
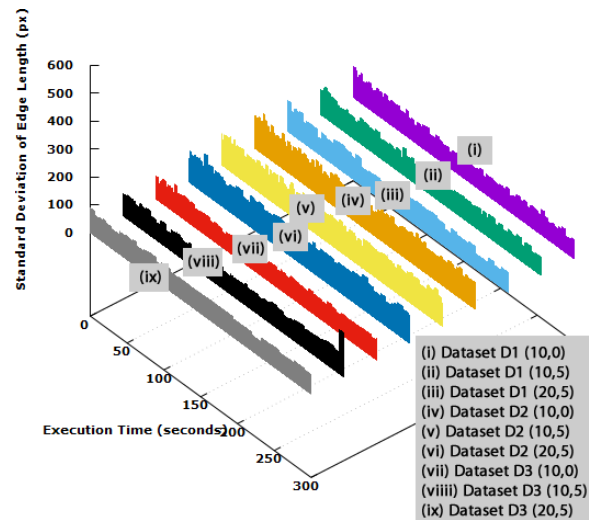
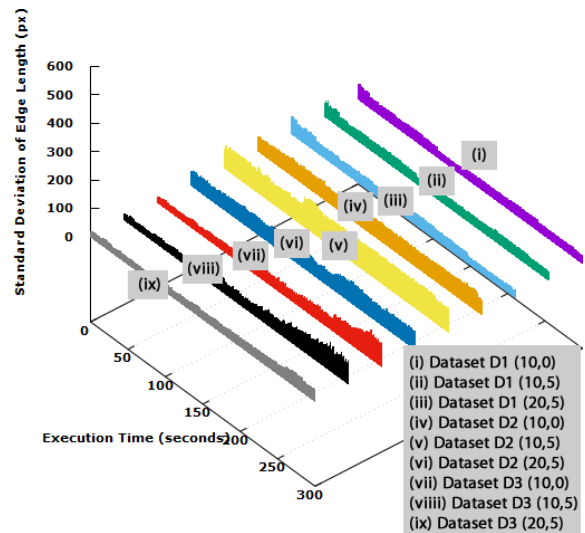

Figure 19: Standard deviation of edge lengths of Gaussian distribution with (a) static FA2, (b) static FR, (c) static KK, (d) static DH, (e) static LinLog algorithms.

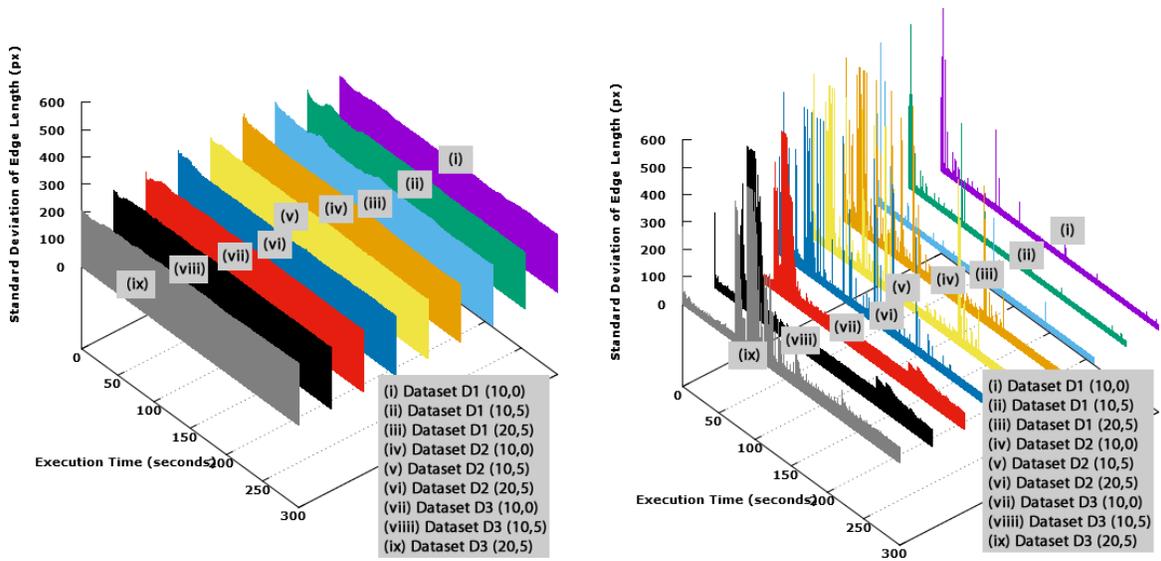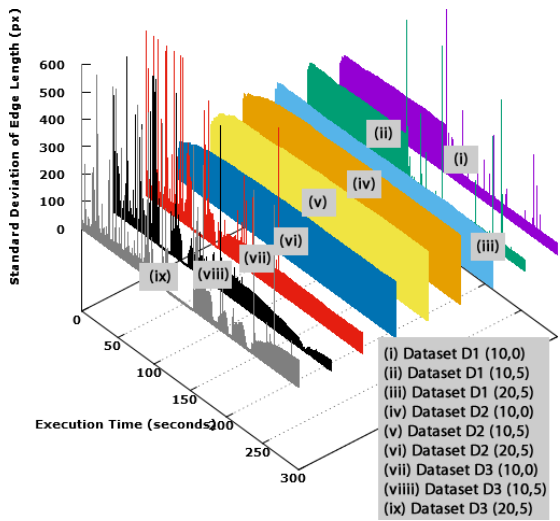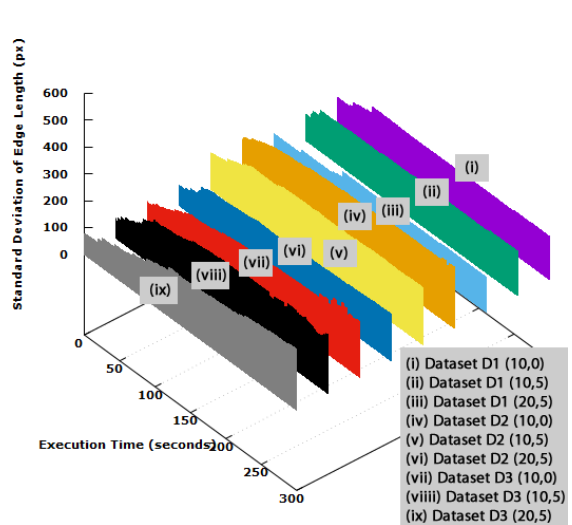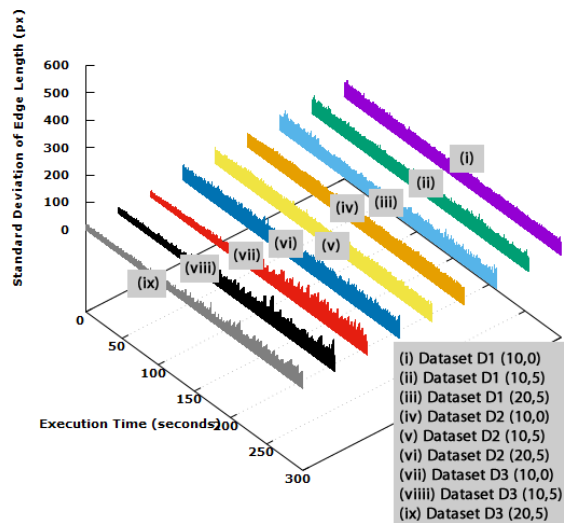

Figure 20: Standard deviation of edge lengths of Gaussian distribution with (a) online FA2, (b) online FR, (c) online KK, (d) online DH, (e) online LinLog algorithms.

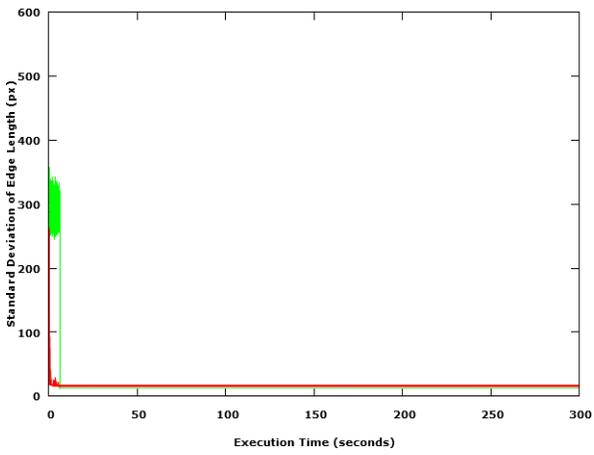
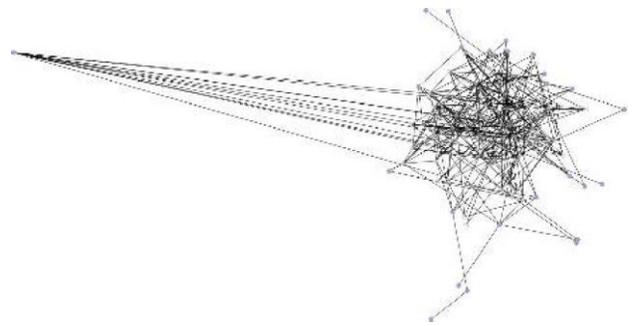

(a)                                                                  (b)

Figure 21: (a) Standard deviation of edge length drops in static KK algorithm, (b) the layout drawing of static KK algorithm with long edge lengths.

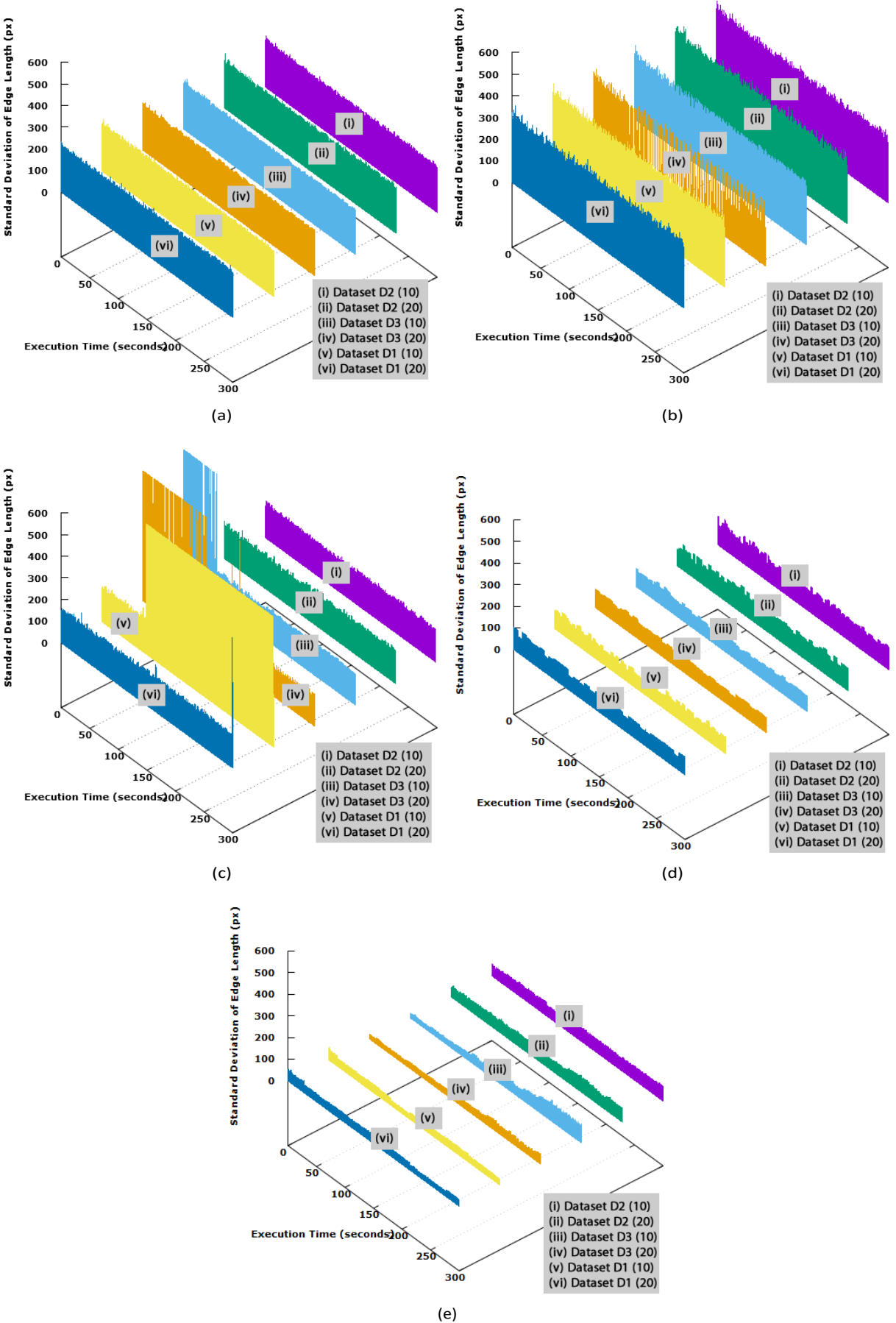

Figure 22: Standard deviation of edge lengths of Poisson distribution with (a) static FA2, (b) static FR, (c) static KK, (d) static DH, (e) static LinLog algorithms.

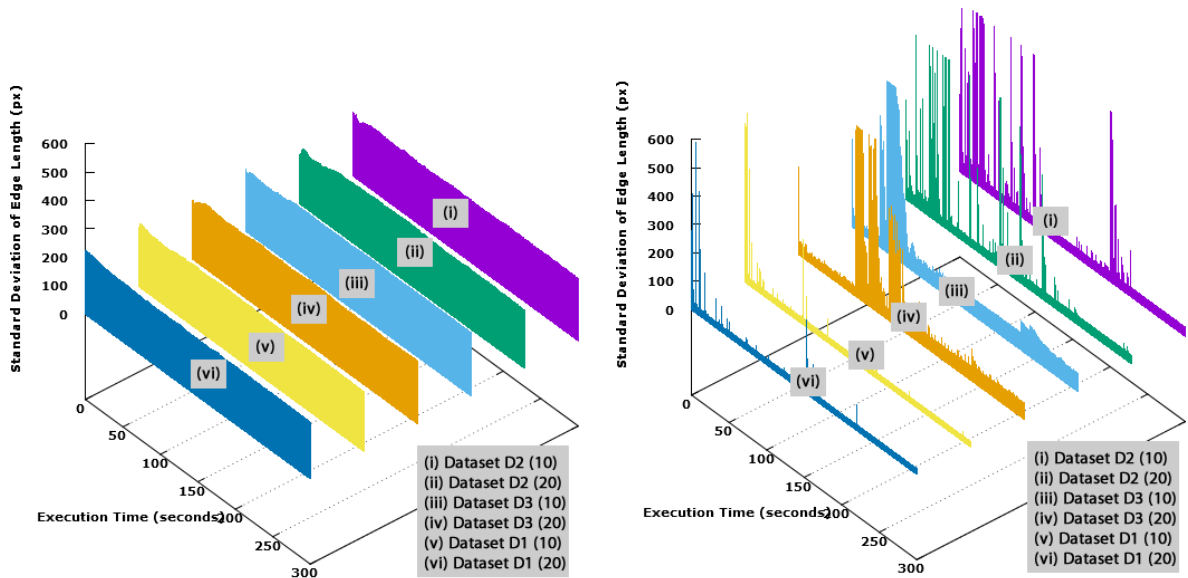
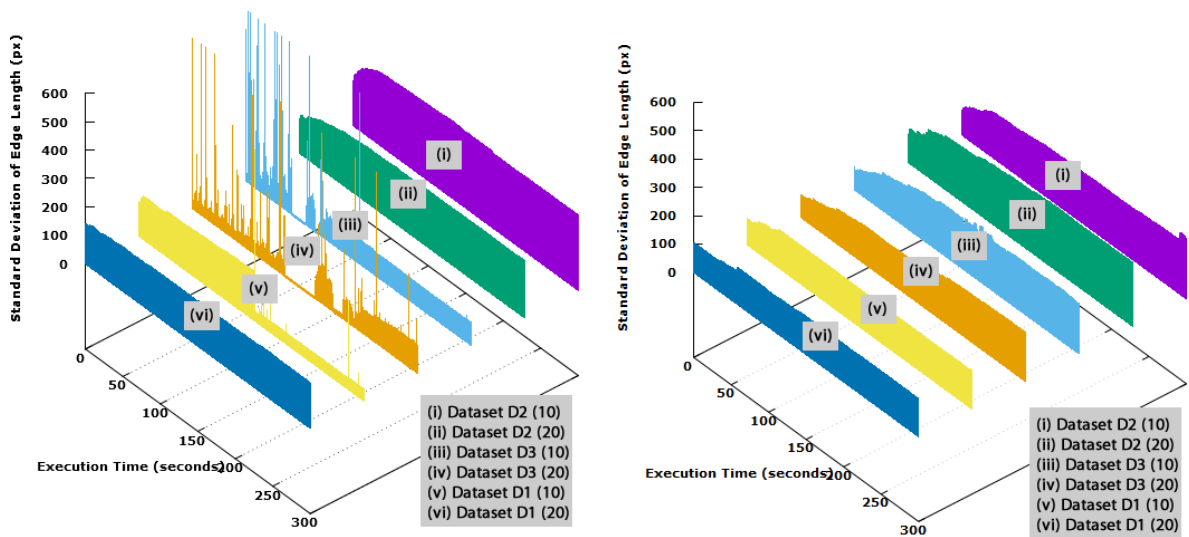
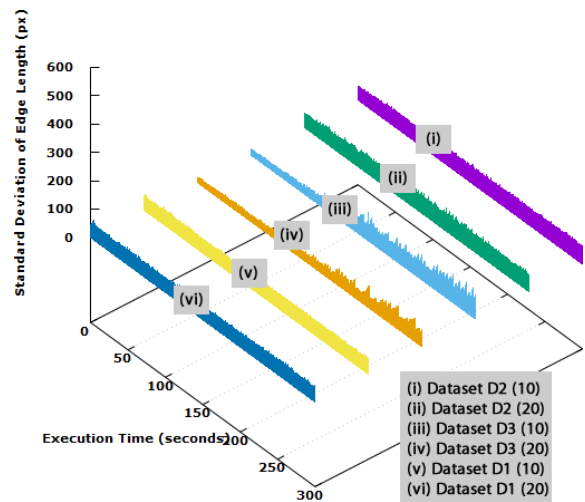

Figure 23: Standard deviation of edge lengths of Poisson distribution with (a) online FA2, (b) online FR, (c) online KK, (d) online DH, (e) online LinLog algorithms.

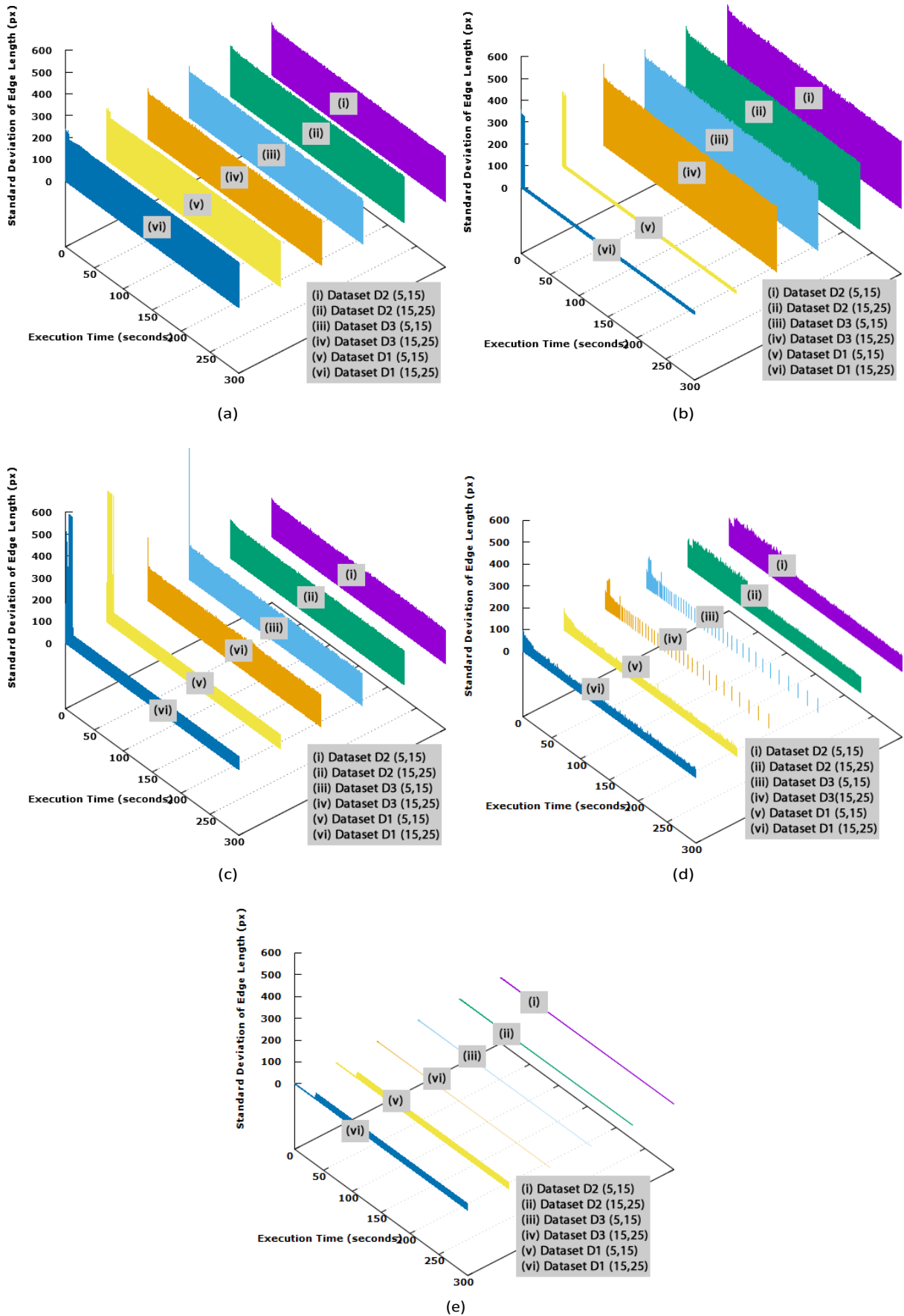

Figure 24: Standard deviation of edge lengths of random distribution with (a) static FA2, (b) static FR, (c) static KK, (d) static DH, (e) static LinLog algorithms.

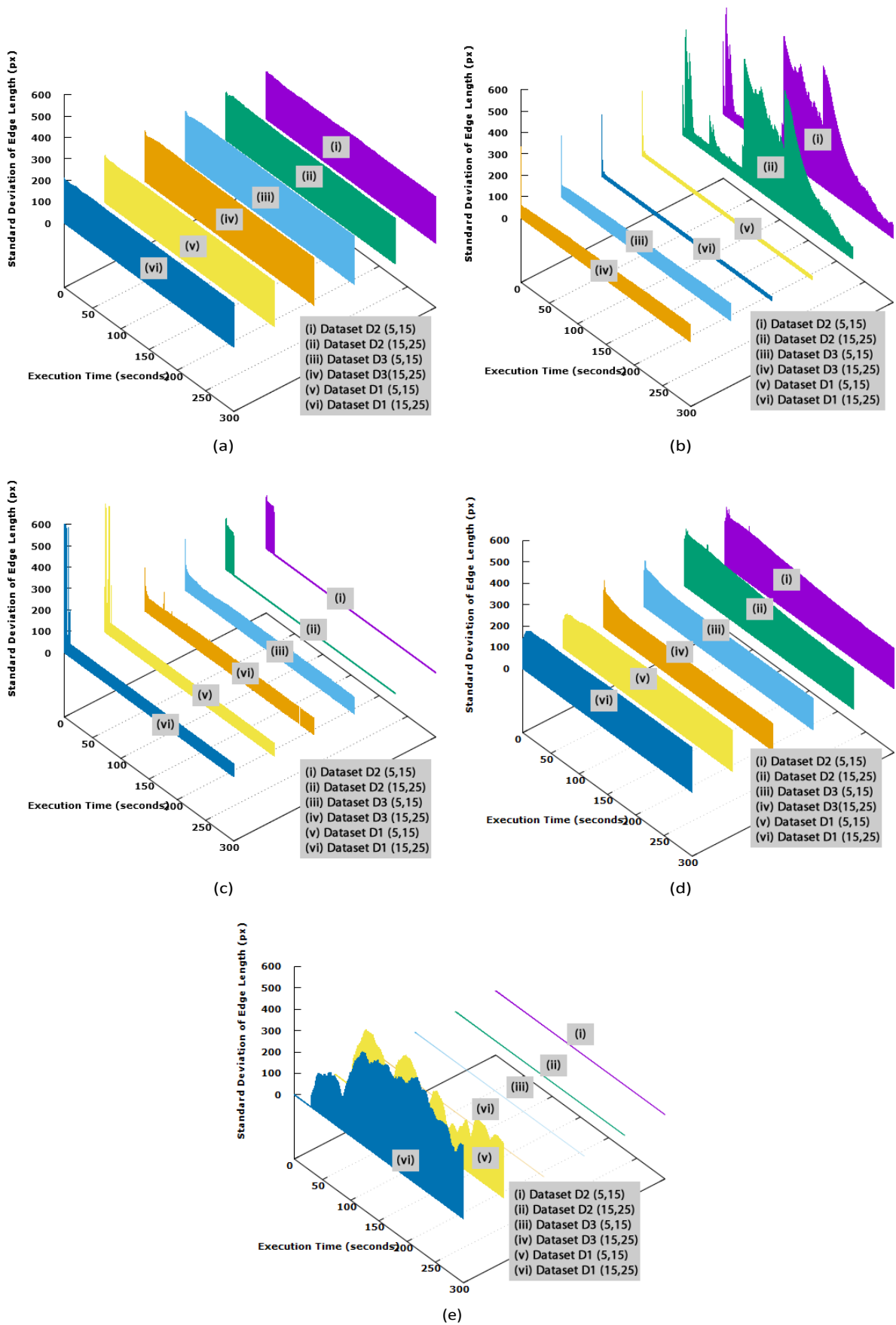

Figure 25: Standard deviation of edge lengths of random distribution with (a) online FA2, (b) online FR, (c) online KK, (d) online DH, (e) online LinLog algorithms.